\documentclass[10pt]{article}
\usepackage{amssymb}
\usepackage{amsmath}
\usepackage{epsfig}

\begin{document}

\title{Tools for network dynamics\thanks{%
Supported by Funda\c{c}\~{a}o para a Ci\^{e}ncia e Tecnologia}}
\author{R. Vilela Mendes\thanks{%
e-mail: vilela@cii.fc.ul.pt} \\
{\small Universidade T\'{e}cnica de Lisboa and Grupo de F\'{i}sica
Matem\'{a}tica,}\\
{\small Complexo Interdisciplinar, Av. Gama Pinto 2, 1649-003 Lisboa,
Portugal}}
\date{}
\maketitle
\tableofcontents

\begin{abstract}
Networks have been studied mainly by statistical methods which emphasize
their topological structure. Here one collects some mathematical tools and
results which might be useful to study both the dynamics of agents living on
the network and the networks themselves as evolving dynamical systems. They
include decomposition of differential dynamics, ergodic techniques,
estimates of invariant measures, construction of non-deterministic automata,
logical approaches, etc. A few network examples are discussed as an
application of the dynamical tools.
\end{abstract}

\section{Introduction}

When modeling extended complex systems, the network concept appears quite
often. The metabolic processes of living beings are a network with the
substrates as nodes, linked together whenever they participate in the same
biochemical reaction. Protein-protein as well as gene expression and
regulation are also networks. Social, economic and political networks are
the backbone of human society, the internet is a network, etc.[Albert \&
Barab\'{a}si, 2002] [Dorogovtsev \& Mendes, 2003]. Most studies deal with
networks as statistical objects, with extensive use of the tools of
statistical mechanics [Pastor-Satorras \textit{et al.}, 2003]. Much less
attention has been paid to the dynamical phenomena taking place in the
networks or to the behavior of the evolving networks as dynamical systems.

For several decades an intensive effort has been dedicated to the study of
low-dimensional dynamical systems, leading to an extensive body of rigorous
results [Katok \& Hasselblatt, 1995]. This exploration is still proceedings
at a good pace with exciting new results, for example, in the dimension
theory of dynamical systems [Pesin, 1997] [Barreira, 2002] and non-uniform
hyperbolicity [Bonatti \textit{et al.}, 2003]. However, the main challenge
for physical applications lies on extended systems, in particular on the
understanding of the dynamics leading from microscopic laws to global
patterns [Cross \& Hohenberg, 1993]. A large amount of numerical work has
been done on the dynamics of these systems which led to the identification
and classification of typical patterns, spatio-temporal chaos, statistical
properties and multistability [Kaneko, 1993] [Kaneko \& Tsuda, 2000]
[Boccaletti \textit{et al.}, 2001] [Boldrighini \textit{et al.}, 2000].
Rigorous results are few, except for regular coupled map lattices
[Bunimovich \& Sinai, 1988] [Coutinho \& Fernandez, 1997a, 1997b] [Jiang \&
Pesin, 1998] [Afraimovich \& Fernandez, 2000] [Gielis \& MacKay, 2000]
[Fernandez \& Guiraud, 2004].

For non-uniform coupling structures there are much less results, statistical
mechanics tools being used mostly to characterize the topological features
of static and evolving networks. Of course the topological structure of the
network is very important for the dynamics. Form affects function and
topology controls the rate at which information or diseases propagate [Boots
\& Sasaki, 1999] [Keeling, 1999] [Pastor-Satorras \& Vespignani, 2002]
[Lloyd \& May, 2001], the robustness under attack and failure [Albert 
\textit{et al.}, 2000] or the adaptation and learning processes that take
place [Ara\'{u}jo \& Vilela Mendes, 2000].

The main purpose of this review is to provide a toolkit for the treatment of
networks (both regular and irregular) as dynamical systems. Results from
differential dynamics and ergodic theory will be presented. To deal with a
network as a dynamical system, three main problems have to be addressed.
First, how to characterize the dynamical behavior of the ``agents'' sitting
on the nodes with interactions defined by the link structure of the network.
Second, if the network topology is not fixed either because the links are
changing in time or because the network itself is growing, to characterize
the network evolution as a dynamical system. Finally to characterize the
interplay between the network topology and the nature of the dynamics.

Section 2 concerns the description of network dynamics by global functions,
limit cycles and multistability, with some examples illustrating both
dynamics on a network and network evolution as a dynamical system. Section 3
discusses ergodic techniques (old and new) and their role in networks.
Section 4 discusses the logic approach to network dynamics. On the important
problem of topology versus dynamics, I will touch only briefly in the
ergodic section, referring for a particular detailed example to [Ara\'{u}jo 
\textit{et al.}, 2003].

In almost all cases the reader should refer to the original papers for
proofs and further developments. Occasionally, as for example in the family
of ergodic parameters or in the relation between synchronization and
Lyapunov spectrum, somewhat more detail is given to the exposition. This
does not mean in any way that I consider this to be a more important topic
than the other interesting results of many authors that are mentioned. If
more detail is included it is because I believe it to be a new result not
included before in any other publication.

\section{Differential dynamics tools}

\subsection{Describing dynamics by global functions}

In many networks, the node dynamics may be modelled by ordinary differential
equations of the form 
\begin{equation}
\frac{dx_{i}}{dt}=X_{i}\left( x\right) =\alpha _{i}+\sum_{j\neq
i}W_{ij}f\left( x_{j}\right) -\gamma _{i}x_{i}  \label{1.1a}
\end{equation}
For a neural network, the $x_{i}^{\prime }s$ might be firing rates and the $%
W_{ij}^{\prime }s$ synaptic intensities [Grossberg, 1988], for a genetic
regulatory system [Tyson \& Ohtmer, 1978] [de Jong, 2002] the variables $%
x_{i}$ would code for the concentrations of RNA, proteins or other metabolic
components and $W_{ij}$ for the production constants (measuring the strength
of $j$ on $i$), $f\left( \cdot \right) $ being the regulation function and $%
-\gamma _{i}x_{i}$ a degradation term, etc.

\subsubsection{Symmetric systems}

Eq.(\ref{1.1a}) is a particular case of the Cohen-Grossberg form [Cohen \&
Grossberg, 1983], used by these authors to describe continuous-time neural
networks, 
\begin{equation}
\frac{dx_{i}}{dt}=a_{i}(x_{i})\left\{
b_{i}(x_{i})-\sum_{j=1}^{n}W_{ij}f_{j}(x_{j})\right\}   \label{1.1}
\end{equation}
Cohen and Grossberg proved that, for the symmetric case ($W_{ij}=W_{ji}$),
the following function 
\begin{equation}
V\left( x_{i}\right) =-\sum_{i=1}^{n}\int^{x_{i}}b_{i}(\xi
_{i})f_{i}^{^{\prime }}(\xi _{i})d\xi _{i}+\frac{1}{2}%
\sum_{j,k=1}^{n}W_{jk}f_{j}(x_{j})f_{k}(x_{k})  \label{1.1b}
\end{equation}
is a Lyapunov function, that is 
\begin{equation}
\frac{d}{dt}V\left( x_{i}\right) \leq 0  \label{1.1c}
\end{equation}
along the orbits if $a_{i}(x_{i})f_{i}^{^{^{\prime }}}(x_{i})>0$. Hopfield's
[Hopfield, 1984] ``energy'' function is a particular case of this result.

The existence of a Lyapunov function is a useful device to characterize the
asymptotically stable states of the network or for the synthesis of networks
with a desired number of stable asymptotic solutions [Cohen, 1992].

In the case of symmetric connections the continuous-time result of Cohen and
Grossberg has been extended to a class of discrete-time systems ([Fogelman
Souli\'{e} \textit{et al.}, 1989] and references therein). For non-symmetric
connections of particular form, namely 
\begin{equation}
\mu _{j}W_{ij}=\mu _{i}W_{ji}  \label{1.1d}
\end{equation}
$\mu _{i}>0$, and time evolution of the connection strengths of Hebbian type 
\begin{equation}
\frac{d}{dt}W_{ij}=-\gamma _{ij}W_{ij}+f_{i}\left( x_{i}\right) f_{j}\left(
x_{j}\right)   \label{1.1e}
\end{equation}
in Sch\"{u}rmann [1989] or $W_{ij}W_{ji}>0$ and $\prod_{C}W_{ij}=%
\prod_{C}W_{ji}$ along every cycle in Fiedler \& Gedeon [1998], Lyapunov
functions may also be constructed.

\subsubsection{General systems}

The Cohen-Grossberg result has been generalized for arbitrary $%
w_{ij}^{\prime }s$ in Vilela Mendes \& Duarte [1992], namely given 
\begin{equation}
\begin{array}{ccl}
W_{ij} & = & W_{ij}^{(S)}+W_{ij}^{(A)} \\ 
W_{ij}^{(S)} & = & \frac{1}{2}\left( W_{ij}+W_{ji}\right)  \\ 
W_{ij}^{(A)} & = & \frac{1}{2}\left( W_{ij}-W_{ji}\right)  \\ 
V^{(S)} & = & -\sum_{i=1}^{n}\int^{x_{i}}b_{i}(\xi _{i})f_{i}^{^{\prime
}}(\xi _{i})d\xi _{i}+\frac{1}{2}%
\sum_{j,k=1}^{n}W_{jk}^{(S)}f_{j}(x_{j})f_{k}(x_{k}) \\ 
H & = & \sum_{i=1}^{n}\int^{x_{i}}\frac{f_{i}(\xi _{i})}{a_{i}(\xi _{i})}%
d\xi _{i}
\end{array}
\label{1.2}
\end{equation}
one has the following

\textit{Theorem} [Vilela Mendes \& Duarte, 1992] If $%
a_{i}(x_{i})/f_{i}^{^{^{\prime }}}(x_{i})>0$ $\forall x,i$ and the matrix $%
W_{ij}^{(A)}$ has an inverse, the vector field $\stackrel{\bullet }{x_{i}}$
in Eq.(\ref{1.1}) decomposes into one gradient and one Hamiltonian
component, $\stackrel{\bullet }{x_{i}}=\stackrel{\bullet }{x_{i}}^{(G)}+%
\stackrel{\bullet }{x_{i}}^{(H)}$, with 
\begin{equation}
\begin{array}{ccl}
\stackrel{\bullet }{x_{i}}^{(G)} & = & -\frac{a_{i}(x_{i})}{f_{i}^{^{\prime
}}(x_{i})}\frac{\partial V^{(S)}}{\partial x_{i}}=-\sum_{j}g_{ij}(x)\frac{%
\partial V^{(S)}}{\partial x_{j}} \\ 
\stackrel{\bullet }{x_{i}}^{(H)} & = & 
-\sum_{j}a_{i}(x_{i})w_{ij}^{(A)}(x)a_{j}(x_{j})\frac{\partial H}{\partial
x_{j}}=\sum_{j}\Gamma _{ij}(x)\frac{\partial H}{\partial x_{j}}
\end{array}
\label{1.3}
\end{equation}
and 
\begin{equation}
\begin{array}{ccl}
g_{ij}(x) & = & \frac{a_{i}(x_{i})}{f_{i}^{^{\prime }}(x_{i})}\delta _{ij}
\\ 
\omega _{ij}(x) & = & -a_{i}(x_{i})^{-1}\left( W^{(A)-1}\right)
_{ij}(x)a_{j}(x_{j})^{-1}
\end{array}
\label{1.4}
\end{equation}
$\left( \sum_{j}\Gamma _{ij}\omega _{jk}=\delta _{ik}\right) $. $g_{ij}(x)$
and $\omega _{jk}(x)$ are the components of the Riemannian metric and the
symplectic form.

\begin{center}
---------
\end{center}

The conditions on $a_{i}(x_{i})$, $f_{i}^{^{^{\prime }}}(x_{i})$ and $%
w_{ij}^{(A)}$ insure that $g$ is a well defined metric and that $\omega $ is
non-degenerate.

The decomposition (\ref{1.3}) is useful, for example, on the design of
oscillatory networks and on the study of gated learning rules [Howse \textit{%
et al.}, 1996]. The nature of the dynamics in the network will depend on the
relative strength of the gradient and the Hamiltonian components. Howse 
\textit{et al.} [1996] propose to measure this relative strength by
comparing $\frac{dV^{(S)}}{dt}$ with $\frac{dH}{dt}$. However, these
quantities vary in space and time and it is the compensation of the two
effects that in particular regions of phase space lead to the attractors of
the dynamics, for example to limit cycles (see below).

The identification, in the differential system (\ref{1.1}), of just one
gradient and one Hamiltonian component, with explicitly known potential and
Hamiltonian functions, is a considerable simplification as compared to a
generic dynamical system. For a general dynamical system a representation by
one or two functions is possible only locally [Vilela Mendes \& Duarte,
1981] and explicit forms for the functions are not easy to obtain [Abarbanel
\& Rouhi, 1987] [Crehan, 1994]. Global decomposition for general dynamical
systems require one gradient and $n-1$ Hamiltonian components [Vilela Mendes
\& Duarte, 1981], namely 
\begin{equation}
\stackrel{\bullet }{x_{i}}=-\sum_{j}g_{ij}(x)\frac{\partial V}{\partial x_{j}%
}+\sum_{k=1}^{n-1}\sum_{j=1}^{n}\left( \omega _{(k)}^{-1}(x)\right) _{ij}%
\frac{\partial H^{(k)}}{\partial x_{j}}  \label{1.5}
\end{equation}
$\left\{ \omega _{(k)}(x)\right\} $ being a set of canonical symplectic
forms adapted to each Hamiltonian component. This result is a generalization
to $n$ dimensions of the 2-dimensional result of Roels [1974]. The first
term in (\ref{1.5}) is the dissipative component and the second one
corresponds to a volume-preserving dynamical system.

The above results lead to a convenient characterization of dynamical systems
of type (\ref{1.1a}) or (\ref{1.1}). For the symmetric case the existence of
a Lyapunov function guarantees global asymptotic stability of the dynamics.
However not all vector fields with a Lyapunov function are differentially
equivalent to a gradient field. Therefore the fact that a gradient vector is
actually obtained gives additional information, namely about structural
stability of the model. A necessary condition for structural stability of
the gradient vector field is the non-degeneracy of the critical points of $%
V^{(S)}$, namely $\det \left\| \frac{\partial ^{2}V^{(S)}}{\partial
x_{i}\partial x_{j}}\right\| \neq 0$ at the points where $\frac{\partial
V^{(S)}}{\partial x_{i}}=0$. In a gradient flow all orbits approach the
critical points as $t\rightarrow \infty $. If the critical points are
non-degenerate, the gradient flow satisfies the conditions defining a
Morse-Smale field, except perhaps the transversality conditions for stable
and unstable manifolds of the critical points. However because Morse-Smale
fields are open and dense in the set of gradient vector fields, any gradient
flow with non-degenerate critical points may always be C$^{1}$-approximated
by a (structurally stable) Morse-Smale gradient field. Therefore given a
symmetric model of the type (\ref{1.1}), the identification of its gradient
nature provides a easy way to check its robustness as a physical model.

Although Lyapunov functions may in some cases be constructed for
discrete-time systems [Fogelman Souli\'{e} \textit{et al.},1989], the
natural functional representation of maps is through generating functions.
This is well known for canonical maps of symplectic manifolds [Amiet \&
Huguenin, 1980] and has been generalized in [Vilela Mendes, 1986] for
noncanonical maps.

The representation of network dynamics by global function applies to neural
networks of several types [Grossberg, 1988], to more general networks [Chua,
1988a, 1988b] and, in view of an established correspondence [Doyne Farmer,
1990], to a large range of connectionistic systems.

\subsection{Cycles}

Existence of limit cycle oscillations in networks is an important issue [Gouz%
\'{e}, 1998] [Plahte \textit{et al.}, 1995] [Snoussi, 1998]. The
decomposition theorems provide a tool to look for candidate orbits with
limit cycle properties. Many years ago Pontryagin [1934], studying small
perturbations of Hamiltonian fields on the plane 
\begin{equation}
\stackrel{\bullet }{x}=\frac{\partial H}{\partial y}+\varepsilon A\left(
x,y,\varepsilon \right) ,\qquad \stackrel{\bullet }{y}=-\frac{\partial H}{%
\partial x}+\varepsilon B\left( x,y,\varepsilon \right)   \label{1.6}
\end{equation}
introduced the notion of \textit{generating cycle} $\gamma \left( c\right) $%
, lying on a level curve $H=c$, when the perturbed equation has a cycle that
depends continuously on $\varepsilon $, for small $\left| \varepsilon
\right| $, and tends to $\gamma \left( c\right) $ when $\varepsilon
\rightarrow 0$. Pontryagin's result states that if $\gamma \left( c\right) $
is a generating cycle, then 
\begin{equation}
I\left( c\right) =\int_{\gamma \left( c\right) }\left( Bdx-Ady\right) =0
\label{1.7}
\end{equation}
the integration being along $\gamma \left( c\right) $ at $\varepsilon =0$.

Further results on the existence of cycles were later proved both for weakly
coupled oscillators and for more general systems with parametrized families
of solutions (see [Hoppensteadt \& Izhikevich, 1997], chapter 9 and
references therein). A generalization of Pontryagin's result to dynamical
systems with constants of motion [Duarte \& Vilela Mendes, 1983], leads to a
necessary condition for the existence of a cycle using the decomposition in (%
\ref{1.5}), namely 
\begin{equation}
\int \left\{ \left( \nabla H^{(i)}\cdot \nabla V\right) +\sum_{k\neq
i}\omega _{(k)}\left( \nabla H^{(i)}\cdot \nabla H^{(k)}\right) \right\}
d\gamma _{i}=0  \label{1.8}
\end{equation}
the integration being along a closed level curve $\gamma _{i}$ of $H_{i}$.

A similar result holds for discrete-time maps which belong to a
differentiable arc with constants of motion [Vilela Mendes \& Duarte, 1982].
A \textit{constant of motion} for a map $f$ defined on a manifold $M$ is a
differentiable function $\Phi :M\rightarrow R$ such that for some orbit $%
\gamma $, $\Phi \circ \gamma =$constant. It generalizes the notion of first
integral which would require this to hold for all orbits. A family of maps $%
f_{\varepsilon }$ is called a \textit{differentiable arc with constants of
motion} if (i) each $f_{\varepsilon }$ has a constant of motion $\Phi
_{\varepsilon }$ for some orbit $\gamma _{\varepsilon }$ ; (ii) The constant
of motion $\Phi _{0}$ of $f_{0}$ is a first integral in a neighborhood of $%
\gamma _{0}$ ; (iii) the maps $\varepsilon \rightarrow f_{\varepsilon
},\varepsilon \rightarrow \gamma _{\varepsilon },\varepsilon \rightarrow
\Phi _{\varepsilon }$ are differentiable. Then 
\begin{equation}
\sum_{n=0}^{N_{0}-1}\left. D\Phi _{0}\left( \gamma _{0}\left( n+1\right)
\right) \right| f^{^{\prime }}\left( \gamma _{0}\left( n\right) \right) =0
\label{1.8a}
\end{equation}
$N_{0}$ being the period of the orbit $\gamma _{0}$.

Both (\ref{1.8}) and (\ref{1.8a}) give only necessary equations for the
existence of limit cycles in the composite dynamics. Nevertheless they are
useful tools to identify limit cycle candidates. Sufficient conditions may
also be obtained in particular low-dimensional cases [Vilela Mendes, 1988,
2000a].

In the same way as the Hamiltonian components of the dynamics provide a tool
to look for limit cycles, the stationary points of the gradient potential
provide information on the multistability of the dynamics and the nature of
their basins of attraction. It is also a tool for the construction of the
invariant measures of the dynamics (see below).

\subsection{Multistability}

Existence of multiple stable states with distinct basins of attraction plays
a significant role in the dynamics of networks, for example in those
associated to the basic processes of life. A genetic regulatory network with
different stable patterns of gene activation explains the emergence of
different phenotypic expressions in the absence of genetic differences
[Laurent \& Kellershohn, 1999]. Examples are also found in population
dynamics [Henson, 2000], neural dynamics [Skarda \& Freeman, 1987] [Freeman,
1992], geophysics, etc. [Vilela Mendes, 2000a]. Extensive numerical work has
been done on classifying different patterns of multistability and their
basins of attraction (see for example [Wuensche, 2002]). Here, I will
concentrate on the dynamical mechanisms leading to the existence of multiple
attractors. Although some of these results are only rigorously proven for
low dimensional systems, their relevance for high dimensional systems is to
be expected.

(i) In the particular case of networks with symmetric connections the
attracting critical points of the potential function $V^{(S)}$ are stable
asymptotic states of the dynamics.

(ii) \textbf{Homoclinic tangencies} (the Newhouse phenomenon)

Contrary to earlier conjectures that generic systems might only have
finitely many attractors, Newhouse [1970, 1974, 1979] proved that a class of
diffeomorphisms in a two-dimensional manifold has infinitely many attracting
periodic orbits (sinks), a result that was later extended to higher
dimensions [Palis \& Viana, 1994]. For two-dimensional manifolds the result
is:

\textbf{Theorem }(Newhouse, Robinson [1983]) Let $f_{\mu }$ be a $C^{3}$ map
in a 2-dimensional manifold with $C^{1}$ dependence on $\mu $ and $\left|
\det \left( T_{a}f_{\mu _{0}}^{n}\right) \right| <1$ and let the
non-degenerate homoclinic tangency be crossed at non-zero speed at $\mu =\mu
_{0}$. Then for $\forall \varepsilon >0$, $\exists (\mu _{1},\mu
_{2})\subset (\mu _{0},\mu _{0}+\varepsilon )$ and a residual subset $%
J\subset (\mu _{1},\mu _{2})$ such that for $\mu \in J$ , $f_{\mu }$ has
infinitely many sinks.

Models of such diffeomorphisms were constructed by Gambaudo and Tresser
[1983] and Wang [1990] proved that the Newhouse set has positive Hausdorff
dimension. After these results, intense research followed on the unfolding
of homoclinic tangencies and an essential question was whether, in addition
to infinitely many sinks, there would also be infinitely many strange
attractors near the homoclinic tangencies. The question was positively
answered by Colli [1998]. The main result is:

\textbf{Theorem }(Colli) Let $f_{0}\in Diff^{\infty }(M)$ be such that $%
f_{0} $ has a homoclinic tangency between the stable and unstable manifolds
of a dissipative hyperbolic saddle $p_{0}$. Then, there is an open set $%
\Omega \subset Diff^{\infty }(M)$ such that

(a) $f_{0}\in \overline{\Omega }$

(b) there is a dense subset $D\subset \Omega $ such that for all $f\in D$ , $%
f$ exhibits infinitely many coexisting H\'{e}non-like strange attractors.

Having established the existence of infinitely many sinks and infinitely
many strange attractors near homoclinic tangencies, a question of practical
importance is the stability of the phenomenon under small random
perturbations of the deterministic dynamics. It turns out that the answer to
this question is negative. Therefore under small random perturbations only
finitely many physical measures will remain.

\textbf{Theorem }(Ara\'{u}jo [2000]) Let $f:M\rightarrow M$ be a
diffeomorphism of class $C^{r},r>1$, of a compact connected boundaryless
manifold $M$ of finite dimension. If $f=f_{a}$ is a member of a parametric
family under parametric noise of level $\varepsilon >0$, that satisfies the
hypothesis:

There are $K\in N$ and $\xi _{0}>0$ such that, for all $k\geq K$ and $x\in M$

(A) $f^{k}(x,\Delta )\supset B^{k}(x),\xi _{0})$ ;

(B) $f^{k}(x,\nu ^{\infty })<<m$ ;

then there is a finite number of probability measures $\mu _{1},\cdots \mu
_{l}$ in $M$ with the properties

1. $\mu _{1},\cdots \mu _{l}$ are physical absolutely continuous probability
measures;

2. supp$\mu _{i}\cap $supp$\mu _{j}$ for all $1\leq i<j\leq l$ ;

3. for all $x\in M$ there are open sets $V_{1}=V_{1}(x),\cdots
,V_{l}=V_{l}(x)\subset \Delta $ such that

(a) $V_{i}\cap V_{j}=\emptyset $, $1\leq i<j\leq l$ ;

(b) $\nu ^{\infty }\left( \Delta \backslash \left( V_{1}\cup \cdots \cup
V_{l}\right) \right) =0$ ;

(c) for all $1\leq i\leq l$ and $\nu ^{\infty }-$ a.e. $t\in V_{i}$ we have 
\[
\lim_{n\rightarrow \infty }\frac{1}{n}\sum_{j=0}^{n-1}\phi \left(
f^{j}(x,t)\right) =\int \phi d\mu 
\]
for every $\phi \in C\left( M,R\right) $. Moreover the sets $V_{1}(x),\cdots
,V_{l}(x)$ depend continuously on $x\in M$ with respect to the distance $%
d_{\nu }\left( A,B\right) =\nu ^{\infty }\left( A\triangle B\right) $
between $\nu ^{\infty }-$\textnormal{mod} $0$ subsets of $\Delta $.

(iii) \textbf{Small dissipative perturbations of conservative systems}

Conservative systems have a large number of coexisting invariant sets,
namely periodic orbits, invariant tori and cantori. By adding a small amount
of dissipation to a conservative system one finds that some of the invariant
sets become attractors. Of course, not all invariant sets of the
conservative system will survive when the dissipation is added. However, for
sufficiently small dissipation many attractors (mainly periodic orbits) have
been observed in typical systems. Poon \& Grebogi [1995], Feudel \textit{et
al.} [1996] and Feudel \& Grebogi [1997] have extensively studied these
effects in the single and double rotor, the H\'{e}non map and the optical
cavity map. They find a large number of attractors for a small amount of
dissipation, in particular in the double rotor map. The large number of
coexisting stable periodic orbits has a complex interwoven basin of
attraction structure, with the basin boundaries permeating most of the state
space. The chaotic component of the dynamics is in the chaotic saddles
embedded in the basin boundary. The systems are also found to be highly
sensitive to small amounts of noise. The problem of migration between
attractors and their stability in multiple-attractor systems has also been
studied by other authors [Weigel \& Atlee Jackson, 1998] [Kaneko, 1997]
[Dutta \textit{et al.}, 1999].

Rigorous results may be obtained in particular cases using the ideas of
deformation stability [Duarte \& Vilela Mendes, 1983] [Vilela Mendes \&
Duarte, 1982] [Vilela Mendes, 1988] [Lima \& Vilela Mendes, 1988]. For
example, let an $\varepsilon -$family of maps be 
\begin{equation}
\begin{array}{lll}
x^{^{\prime }} & = & bx+y+f(x,y,\varepsilon ) \\ 
y^{^{\prime }} & = & y+g(x,y,\varepsilon )
\end{array}
\label{1.20}
\end{equation}
with $f(x,y,0)=g(x,y,0)=0$ For $\varepsilon =0$ the map has marginally
stable periodic orbits of all periods. Under perturbation some of the orbits
become stable ones.

\textbf{Theorem }[Vilela Mendes, 2000a]\textbf{\ }If $f$ and $g$ are jointly 
$C^{2}$ in $(x,y,\varepsilon )$ with $f(x,y,0)=g(x,y,0)=0$, there is an $%
\overline{\varepsilon }$ such that for $\left| \varepsilon \right| <\left| 
\overline{\varepsilon }\right| $ an interior orbit of period $p$ of the
unperturbed map becomes a stable orbit of the perturbed map if and only if:

(1) $\sum\limits_{n=0}^{p-1}\partial _{\varepsilon
}g(x_{n}^{(0)},y^{(0)},0)\mid _{\varepsilon =0}=0$

(2) $\varepsilon \partial _{\varepsilon }\sum\limits_{n=0}^{p-1}\left\{
\partial _{x}g(x_{n}^{(0)},y^{(0)},\varepsilon )+(1-b)\partial
_{y}g(x_{n}^{(0)},y^{(0)},\varepsilon )\right\} \mid _{\varepsilon =0}<0$

(iv) \textbf{Coupled dynamical systems near period-doubling accumulation
points}

An example is a system of two coupled quadratic maps [Carvalho \textit{et al.%
}, 2001],

\begin{equation}
\begin{array}{c}
x_{1}(t+1)=1-\mu _{*}\left( (1-\varepsilon )x_{1}(t)+\varepsilon
x_{2}(t)\right) ^{2} \\ 
x_{2}(t+1)=1-\mu _{*}\left( \varepsilon x_{1}(t)+(1-\varepsilon
)x_{2}(t)\right) ^{2}
\end{array}
\label{1.21}
\end{equation}
with $x\in [-1,1]$, and $\mu _{*}=1.401155...$ , which is the parameter
value of the period doubling accumulation point. The result is that \textit{%
for any }$N$\textit{\ there is a sufficiently small }$\varepsilon \left(
N\right) $\textit{\ such that there are }$N$\textit{\ distinct stable
periodic orbits}.

\subsection{Network examples}

\subsubsection{An excitatory-inhibitory network}

Excitatory-inhibitory networks exhibit a rich variety of activity patterns.
They have been identified as underlying several biological processes like
image segmentation, sleep rhythms, control of movement in the basal ganglia,
etc. [Terman, 2002]. Here one considers a simple network with two
populations, one composed of excitatory and the other of inhibitory cells.
The equations of motion are 
\begin{equation}
\begin{array}{lllll}
\stackrel{\bullet }{x}_{i} & = & -\alpha \sum_{j=N+1}^{2N}f\left(
x_{j}\right) +\beta g\left( t\right)  &  & i=1,\cdots ,N \\ 
\stackrel{\bullet }{x}_{i} & = & \alpha \sum_{j=1}^{N}f\left( x_{j}\right)
-\gamma x_{i} &  & i=N+1,\cdots ,2N
\end{array}
\label{1.31}
\end{equation}
The first population ($i=1,\cdots ,N$) receives a time-dependent driving
external signal $g\left( t\right) $ and inhibitory inputs from the second
population. The second population receives excitatory inputs from the first
population and has a decay rate $-\gamma $. The activation function is 
\begin{equation}
f\left( x\right) =1-\exp \left( -\mu x\right)   \label{1.32}
\end{equation}
In terms of the global functions described before, the dynamics has a simple
form 
\begin{equation}
\stackrel{\bullet }{x}_{i}=-\frac{\partial V}{\partial x_{i}}+\sum_{j}\Gamma
_{ij}\frac{\partial H}{\partial x_{j}}  \label{1.33}
\end{equation}
with 
\begin{equation}
\begin{array}{lll}
V & = & -\beta g\left( t\right) \sum_{i}x_{i}+\frac{\gamma }{2}%
\sum_{i}x_{i}^{2} \\ 
H & = & \alpha \sum_{i}\int^{x_{i}}f\left( \xi \right) d\xi 
\end{array}
\label{1.34}
\end{equation}
$\Gamma $ being the matrix\footnote{%
the degeneracy of the symplectic form is lifted by a particular choice of
coordinates} 
\begin{equation}
\Gamma =\left( 
\begin{array}{llllllll}
0 & 0 & \cdots  & 0 & -1 & -1 & \cdots  & -1 \\ 
0 & 0 & \cdots  & 0 & -1 & -1 & \cdots  & -1 \\ 
\cdots  & \cdots  & \cdots  & \cdots  & \cdots  & \cdots  & \cdots  & \cdots 
\\ 
0 & 0 & \cdots  & 0 & -1 & -1 & \cdots  & -1 \\ 
1 & 1 & \cdots  & 1 & 0 & 0 & \cdots  & 0 \\ 
1 & 1 & \cdots  & 1 & 0 & 0 & \cdots  & 0 \\ 
\cdots  & \cdots  & \cdots  & \cdots  & \cdots  & \cdots  & \cdots  & \cdots 
\\ 
1 & 1 & \cdots  & 1 & 0 & 0 & \cdots  & 0
\end{array}
\right)   \label{1.35}
\end{equation}

\begin{figure}[htb]
\begin{center}
\psfig{figure=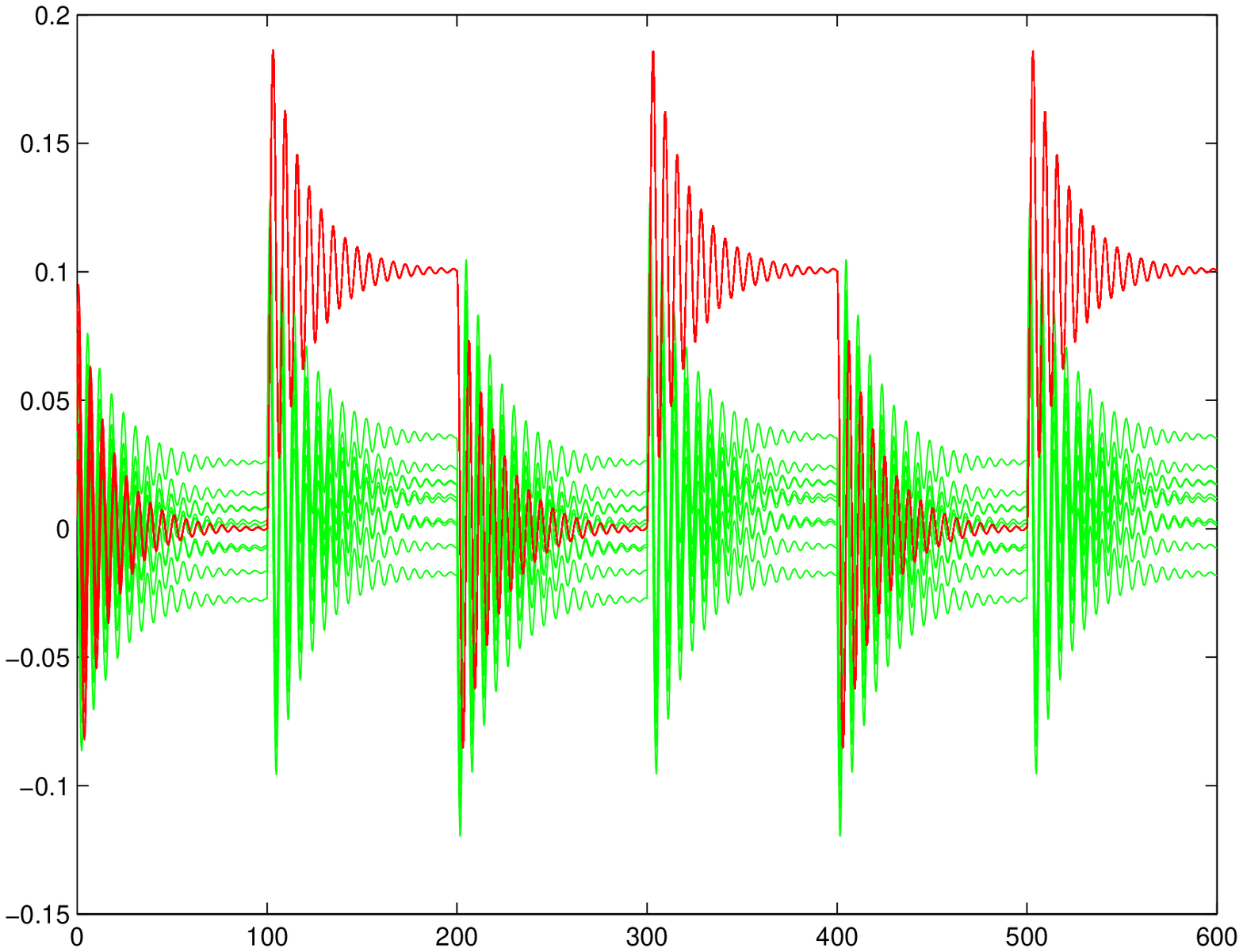,width=9truecm}
\end{center}
\caption{Dynamics of the excitatory-inhibitory network (\ref{1.31}) ($\alpha
=1,\beta =0.1,\gamma =0.1$) for a square-wave signal $g\left( t\right) $}
\end{figure}

\begin{figure}[htb]
\begin{center}
\psfig{figure=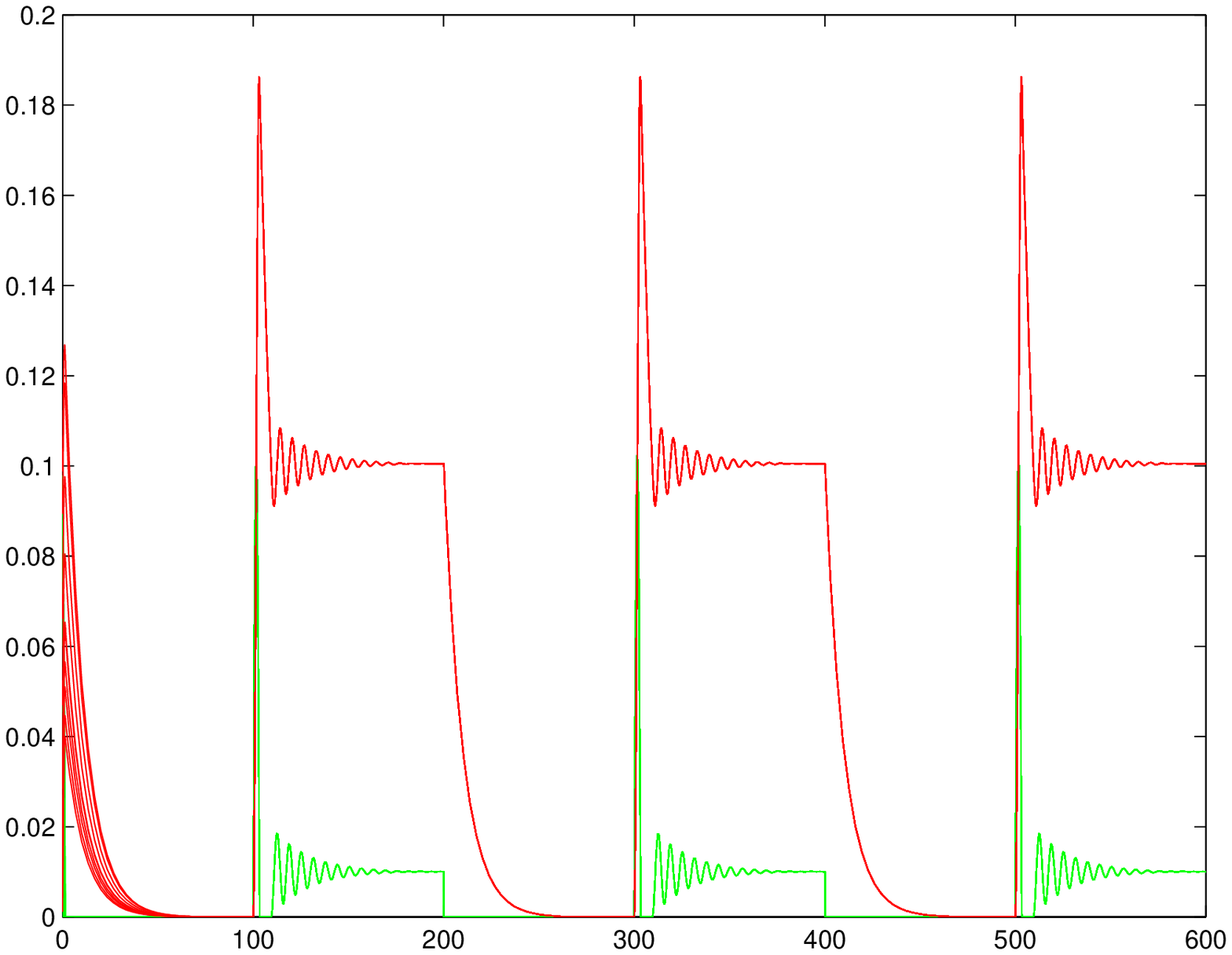,width=9truecm}
\end{center}
\caption{Dynamics of the excitatory-inhibitory network (\ref{1.36}) ($\alpha
=1,\beta =0.1,\gamma =0.1$) for a square-wave signal $g\left( t\right) $}
\end{figure}

Except for the contribution of the driving signal and the decay constant,
the dynamics is Hamiltonian, the symmetric connection coefficients $%
w_{ij}^{(S)}$ in (\ref{1.2}) vanishing identically. Therefore, the reaction
of the network to an external signal is simply a damped oscillation. This is
illustrated in Fig.1 where, starting from a random initial condition and
after a relaxation period, a square wave $g\left( t\right) $ with zero
baseline is applied to the network. The base level of each unit depends on
the initial conditions.

Notice however that, if the agents in the network are biological cells,
their activation cannot be negative. Then, the dynamical system should be
modified from $\stackrel{\bullet }{x}_{i}=F_{i}\left( x\right) $ (where $%
F_{i}\left( x\right) $ is the right hand side of (\ref{1.31})) to 
\begin{equation}
\stackrel{\bullet }{x}_{i}=F_{i}\left( x\right) \cdot \mathnormal{OR}\left( 
\mathnormal{sign}\left( x_{i}\right) ,\mathnormal{sign}\left( F_{i}\right)
\right)  \label{1.36}
\end{equation}
OR being the logical function 
\begin{equation}
\begin{tabular}{lll}
& $+$ & $-$ \\ \cline{2-3}
$+$ & \multicolumn{1}{|l}{1} & \multicolumn{1}{|l|}{1} \\ 
\cline{2-3}\cline{3-3}
$-$ & \multicolumn{1}{|l}{1} & \multicolumn{1}{|l|}{0} \\ \cline{2-3}
\end{tabular}
\label{1.37}
\end{equation}

In this case, because the right hand side is no longer of Cohen-Grossberg
form, the decomposition (\ref{1.33})$-$(\ref{1.34}) gives only qualitative
information on the dynamical behavior of the system. This is illustrated in
Fig.2 where the same square-wave driving force, as in Fig.1, is applied to
the modified network.

\subsubsection{A gene regulation network}

The p53 gene was one of the first tumour-suppressor genes to be identified,
its protein acting as an inhibitor of uncontrolled cell growth. The p53
protein has been found not to be acting properly in most human cancers, due
either to mutations in the gene or inactivation by viral proteins or
inhibiting interactions with other cell products. Although apparently not
required for normal growth and development, p53 is critical in the
prevention of tumour development, contributing to DNA repair, inhibiting
angiogenesis and growth of abnormal or stressed cells [May \& May, 1999]
[Vogelstein \textit{et al.}, 2000] [Woods \& Vousden, 2001] [Taylor \textit{%
et al.}, 1999] [Vousden, 2000]. In addition to its beneficial anticancer
activities it may also have some detrimental effects in human aging
[Sharpless \& DePinho, 2002].

The p53 gene does not act by itself, but through a very complex network of
interactions [Kohn, 1999]. Here I will discuss a simplified network, which
although not being accurate in biological detail, tends to capture the
essential features of the p53 network as it is known today. In particular,
several different products and biological mechanisms are lumped together
into a single node when their function is identical. The network is depicted
in Fig.3. The arrows and signs denote the excitatory or inhibitory action of
each node on the others and the letters $b,g,c,r,p,m,a$ denote their
intensities (or concentrations).

\begin{figure}[htb]
\begin{center}
\psfig{figure=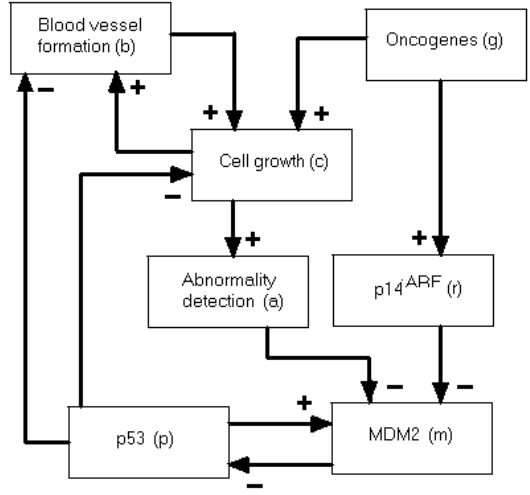,width=9truecm}
\end{center}
\caption{A simplified p53 network model}
\end{figure}

The p53 protein is assumed to be produced at a fixed rate ($k_{p}$) and to
be degraded after ubiquitin labelling. The MDM2 protein being one of the
main enzymes involved in ubiquitin labelling, the inhibitory node ($m$) is
denoted MDM2. There is a positive feedback loop from p53 to MDM2, because
the p53 protein, binding to the regulatory region of the MDM2 gene,
stimulates the transcription of this gene into mRNA.

Under normal circumstances the network is ``off'' or operates at a low
level. The main activation pathways are the detection of cell anomalies ($a$%
), like DNA damage, or aberrant growth signals, such as those resulting from
the expression of several oncogenes (the p14$^{ARF}$ pathway, $r$) . They
inhibit the degradation of the p53 protein, which may then reach a high
level. There are several distinct activation pathways. For example, in some
cases phosphorylation of the p53 protein blocks its interaction with MDM2
and in others it is a protein that binds to MDM2 and inhibits its action.
However, the end result being a decrease in the MDM2 efficiency, they may
both be described as inhibitory inputs to the MDM2 node.

The p53 protein controls cell growth and proliferation, either by blocking
the cell division cycle, or activating apoptosis or inhibiting the
blood-vessel formation ($b$) that is stimulated by several tumors. In our
simplified p53 network all these effects are coded on the following set of
equations

\begin{equation}
\begin{array}{lll}
\frac{dp}{dt} & = & k_{p}-W_{pm}f_{m}\left( m\right) \\ 
\frac{dm}{dt} & = & W_{mp}f_{p}\left( p\right) -W_{mr}f_{r}\left( r\right)
-W_{ma}a-\gamma _{m}^{\prime }m \\ 
\frac{db}{dt} & = & W_{bc}f_{c}\left( c\right) -W_{bp}f_{p}\left( p\right)
\\ 
\frac{dc}{dt} & = & W_{cg}g+W_{cb}f_{b}\left( b\right) -W_{cp}f_{p}\left(
p\right) \\ 
\frac{dr}{dt} & = & W_{rg}g-\gamma _{r}r
\end{array}
\label{1.38}
\end{equation}
One should note that an increased level of cellular p53 is not by itself
sufficient for it to become a transcriptional activator controlling cell
growth. Conformational changes of the protein are also needed which are
stimulated by the activation pathways or may be therapeutically induced.
Also some viruses produce proteins that inactivate p53. All this means that
in reality some of the coupling constants in Eqs.(\ref{1.38}), (for example $%
W_{cp}$) may also be dynamical variables.

\begin{figure}[htb]
\begin{center}
\psfig{figure=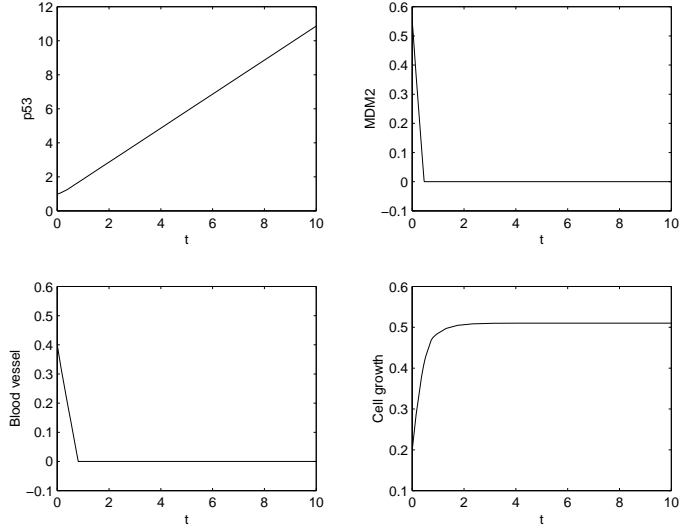,width=9truecm}
\end{center}
\caption{Time evolution of the network (\ref{1.42})}
\end{figure}

\begin{figure}[htb]
\begin{center}
\psfig{figure=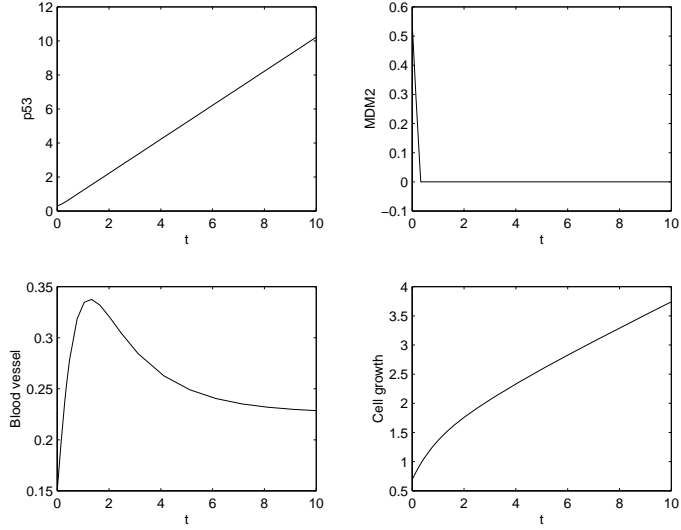,width=9truecm}
\end{center}
\caption{Same as Fig.4 with different initial conditions}
\end{figure}

The regulation functions $f\left( \cdot \right) $ are positive nonlinear
functions with a threshold and a saturation level. By shifting variables to
compensate for thresholds and rescaling the coupling constants they may be
normalized by the coefficient of the linear part, that is 
\begin{equation}
f_{i}\left( x_{i}\right) =x_{i}+\cdots  \label{1.39}
\end{equation}

With a rescaling of $p,m,b,c,r$ and redefinition of the constants we may
consider 
\begin{equation}
k_{p}=W_{mp}=W_{bc}=W_{cg}=W_{rg}=1  \label{1.40}
\end{equation}
Furthermore, from the last equation in (\ref{1.38}) 
\begin{equation}
r\left( t\right) =\frac{1}{\gamma _{r}}g+\left( r\left( 0\right) -\frac{g}{%
\gamma _{r}}\right) e^{-\gamma _{r}r}  \label{1.41}
\end{equation}
Replacing $r$ by its steady state value $g/\gamma _{r}$, and rescaling $%
W_{mr}$ we are left with 
\begin{equation}
\begin{array}{lll}
\frac{dp}{dt} & = & 1-W_{pm}f_{m}\left( m\right) \\ 
\frac{dm}{dt} & = & f_{p}\left( p\right) -W_{mr}g-W_{ma}a-\gamma _{m}m \\ 
\frac{db}{dt} & = & f_{c}\left( c\right) -W_{bp}f_{p}\left( p\right) \\ 
\frac{dc}{dt} & = & g+W_{cb}f_{b}\left( b\right) -W_{cp}f_{p}\left( p\right)
\end{array}
\label{1.42}
\end{equation}
a system of four dynamical variables and two control parameters $g$ and $a$.

Using the dynamics decomposition discussed in Section 2 one obtains 
\begin{equation}
\begin{array}{lll}
V^{(S)} & = & -\int^{p}f_{p}^{^{\prime }}\left( \xi \right) d\xi
-\int^{c}f_{c}^{^{\prime }}\left( \xi \right) d\xi +\int^{m}\left(
W_{mr}g+W_{ma}a+\gamma _{m}\xi \right) f_{m}^{^{\prime }}\left( \xi \right)
d\xi \\ 
&  & +\frac{1}{2}\sum_{x_{i}=p,m,b,c}W_{ij}^{(S)}f_{i}\left( x_{i}\right)
f_{j}\left( x_{j}\right) \\ 
H & = & \sum_{x_{i}=p,m,b,c}\int^{x_{i}}f_{i}\left( \xi \right) d\xi
\end{array}
\label{1.49}
\end{equation}
\begin{equation}
g_{ij}=\frac{1}{f_{i}^{^{\prime }}\left( x_{i}\right) }\delta _{ij}\qquad
\Gamma _{ij}=-W_{ij}^{(A)}  \label{1.50}
\end{equation}
\begin{equation}
\begin{tabular}{|c|c|c|}
\hline
$(i,j)$ & $W_{ij}^{(S)}$ & $W_{ij}^{(A)}$ \\ \hline
$pm$ & $\frac{1}{2}\left( W_{pm}-1\right) $ & $\frac{1}{2}\left(
W_{pm}+1\right) $ \\ \hline
$pb$ & $\frac{1}{2}W_{bp}$ & $-\frac{1}{2}W_{bp}$ \\ \hline
$pc$ & $\frac{1}{2}W_{cp}$ & $-\frac{1}{2}W_{cp}$ \\ \hline
$mb$ & $0$ & $0$ \\ \hline
$mc$ & $0$ & $0$ \\ \hline
$bc$ & $-\frac{1}{2}\left( W_{cb}+1\right) $ & $\frac{1}{2}\left(
W_{cb}-1\right) $ \\ \hline
\end{tabular}
\label{1.51}
\end{equation}
The reaction of $m$ and $c$ to external stimuli ($g$ and $a$) and the
production rate of $p$ are coded on the first three terms of the potential
function $V^{(S)}$. For coupling constants of order unit, one sees from (\ref
{1.51}) the existence of a damped Hamiltonian oscillation for the $p-m$
system, and a dangerous runaway behavior of $b-c$ arising from its
dominantly gradient dynamics. The action of $p$ on $b$ and $c$ is of mixed
gradient-Hamiltonian type. Hence, from inspection of the nature of the
global functions describing the dynamics, one concludes that (at least in
this model) the controlling action of p53 may only be effective in
particular circumstances. That is, it will depend on the initial conditions.
This conclusion is now checked by a detailed study of the solutions.

Consider first the linear approximation to the system. The solutions are,
for the $p-m$ system 
\begin{equation}
\begin{array}{lll}
p\left( t\right) & = & \stackrel{-}{p}+p^{\prime }\left( t\right) \\ 
m\left( t\right) & = & \stackrel{-}{m}+m^{\prime }\left( t\right)
\end{array}
\label{1.53}
\end{equation}
with 
\begin{equation}
\begin{array}{lll}
\stackrel{-}{p} & = & W_{mr}g+W_{ma}a+\frac{\gamma _{m}}{W_{pm}} \\ 
\stackrel{-}{m} & = & \frac{1}{W_{pm}}
\end{array}
\label{1.54}
\end{equation}
\begin{equation}
\begin{array}{ll}
p^{\prime }\left( t\right) = & e^{-\gamma _{m}/2}\left\{ \left( p\left(
0\right) -\stackrel{-}{p}\right) \cos \alpha t+\frac{1}{\alpha }\left( \frac{%
\gamma _{m}}{2}\left( p\left( 0\right) -\stackrel{-}{p}\right) -W_{pm}\left(
m\left( 0\right) -\frac{1}{W_{pm}}\right) \right) \sin \alpha t\right\} \\ 
m^{\prime }\left( t\right) = & e^{-\gamma _{m}/2}\left\{ \left( m\left(
0\right) -\frac{1}{W_{pm}}\right) \cos \alpha t+\frac{1}{\alpha }\left(
p\left( 0\right) -\stackrel{-}{p}-\frac{\gamma _{m}}{2}\left( m\left(
0\right) -\frac{1}{W_{pm}}\right) \right) \sin \alpha t\right\}
\end{array}
\label{1.55}
\end{equation}
and $\alpha =\sqrt{W_{pm}-\gamma ^{2}/4}$.

As expected, one sees a damped oscillatory behavior of the $p-m$ system and,
in the absence of stimuli ($a=g=0$) the $p$ level is small and controlled by
the degradation of $m$.

For $b$ and $c$ one now obtains 
\begin{equation}
\left( 
\begin{array}{l}
b\left( t\right) \\ 
c\left( t\right)
\end{array}
\right) =\left( 
\begin{array}{l}
W_{bp}\stackrel{-}{p} \\ 
\frac{\left( W_{cp}\stackrel{-}{p}-g\right) }{W_{cb}}
\end{array}
\right) -\int_{0}^{t}e^{A\left( t-\tau \right) }\left( 
\begin{array}{l}
W_{bp}p^{\prime }\left( \tau \right) \\ 
W_{cp}p^{\prime }\left( \tau \right)
\end{array}
\right) d\tau +e^{At}\left( 
\begin{array}{l}
c\left( 0\right) -W_{bp}\stackrel{-}{p} \\ 
b\left( 0\right) -\frac{W_{cp}\stackrel{-}{p}-g}{W_{cb}}
\end{array}
\right)  \label{1.56}
\end{equation}
where $A$ is the matrix 
\begin{equation}
A=\left( 
\begin{array}{ll}
0 & 1 \\ 
W_{cb} & 0
\end{array}
\right)  \label{1.57}
\end{equation}

This matrix has eigenvalues $\pm \sqrt{W_{cb}}$ implying that $b\left(
t\right) $ and $c\left( t\right) $ are going to have terms proportional to $%
\exp \left( t\sqrt{W_{cb}}\right) $ and $\exp \left( -t\sqrt{W_{cb}}\right) $%
. Hence $p$ (p53) will only have a controlling effect on cell proliferation
if the coefficient of the exponentially growing terms becomes negative.
Multiplying (\ref{1.56}) on the left by the matrix $\frac{1}{2}\left( 
\begin{array}{ll}
\sqrt{W_{cb}} & 1 \\ 
-\sqrt{W_{cb}} & 0
\end{array}
\right) $ that diagonalizes $A$ one obtains the coefficient of the
exponentially growing term 
\begin{equation}
\begin{array}{lll}
B\left( t\right) & = & e^{t\sqrt{W_{cb}}}\left\{ \frac{\sqrt{W_{cb}}}{2}%
\left( c\left( 0\right) -W_{bp}\stackrel{-}{p}\right) +\frac{1}{2}\left(
b\left( 0\right) -\frac{W_{cp}\stackrel{-}{p}-g}{W_{cb}}\right) \right\} \\ 
&  & -\int_{0}^{t}e^{\left( t-\tau \right) \sqrt{W_{cb}}}\left( \frac{\sqrt{%
W_{cb}}}{2}W_{bp}+\frac{1}{2}W_{cp}\right) p^{\prime }\left( \tau \right)
d\tau
\end{array}
\label{1.59}
\end{equation}

\begin{figure}[htb]
\begin{center}
\psfig{figure=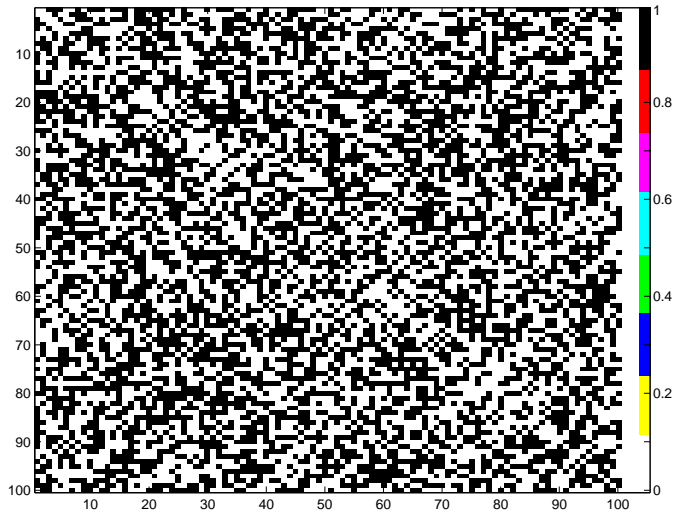,width=9truecm}
\end{center}
\caption{Typical equilibrium configuration of network connections evolved
according to Eqs. (\ref{1.65})-(\ref{1.66}) ($\alpha =1,\beta =0$)}
\end{figure}

\begin{figure}[hbt]
\begin{center}
\psfig{figure=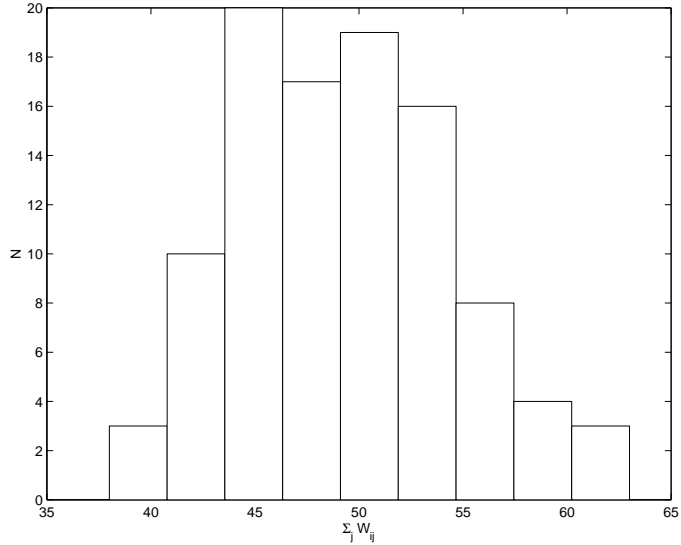,width=9truecm}
\end{center}
\caption{Degree distribution of the network in Fig.6}
\end{figure}

The conclusion is that control of cell proliferation is obtained only if $%
\exists t$ such that $B\left( t\right) <0$. Therefore it depends strongly on
the initial conditions. This conclusion, inferred both from the dynamical
decomposition and the linear approximation is borne out by simulation of the
nonlinear problem. Figs.4 and 5 show two time evolutions of the equations (%
\ref{1.38}) with $f\left( x\right) =\tanh \left( x\right) $ , $%
W_{pm}=W_{mr}=W_{ma}=W_{bp}=W_{cb}=W_{cp}=1$, $\gamma _{m}=0.01$, $g=a=1$
and the vector field $\stackrel{\bullet }{x}_{i}=F_{i}$ truncated to $%
\stackrel{\bullet }{x}_{i}=F_{i}\cdot $OR$\left( \mathnormal{sign}\left(
x_{i}\right) ,\mathnormal{sign}\left( F_{i}\right) \right) $, because
concentrations cannot become negative. The behavior depends strongly on the
value of the initial conditions. In conclusion, the implication (of the
model) is that unless p53 starts acting soon enough its action is useless
and other means have to be used to control cell proliferation.

\subsection{Evolving networks}

In many networks found in Nature, as important as the structure of the
network, is the path that the network took to reach that final state. Social
or economic networks, industrial, transportation and communication networks,
ecological webs, biological networks, all are examples of evolving networks.
In many cases their complex structure is a simple consequence of the
principles of their growth. Several network growth schemes have been studied
(see Albert \& Barab\'{a}si [2002], Dorogovtsev \& Mendes [2003] and
Pastor-Satorras \textit{et al.} [2003] for reviews). 
\begin{figure}[tbh]
\begin{center}
\psfig{figure=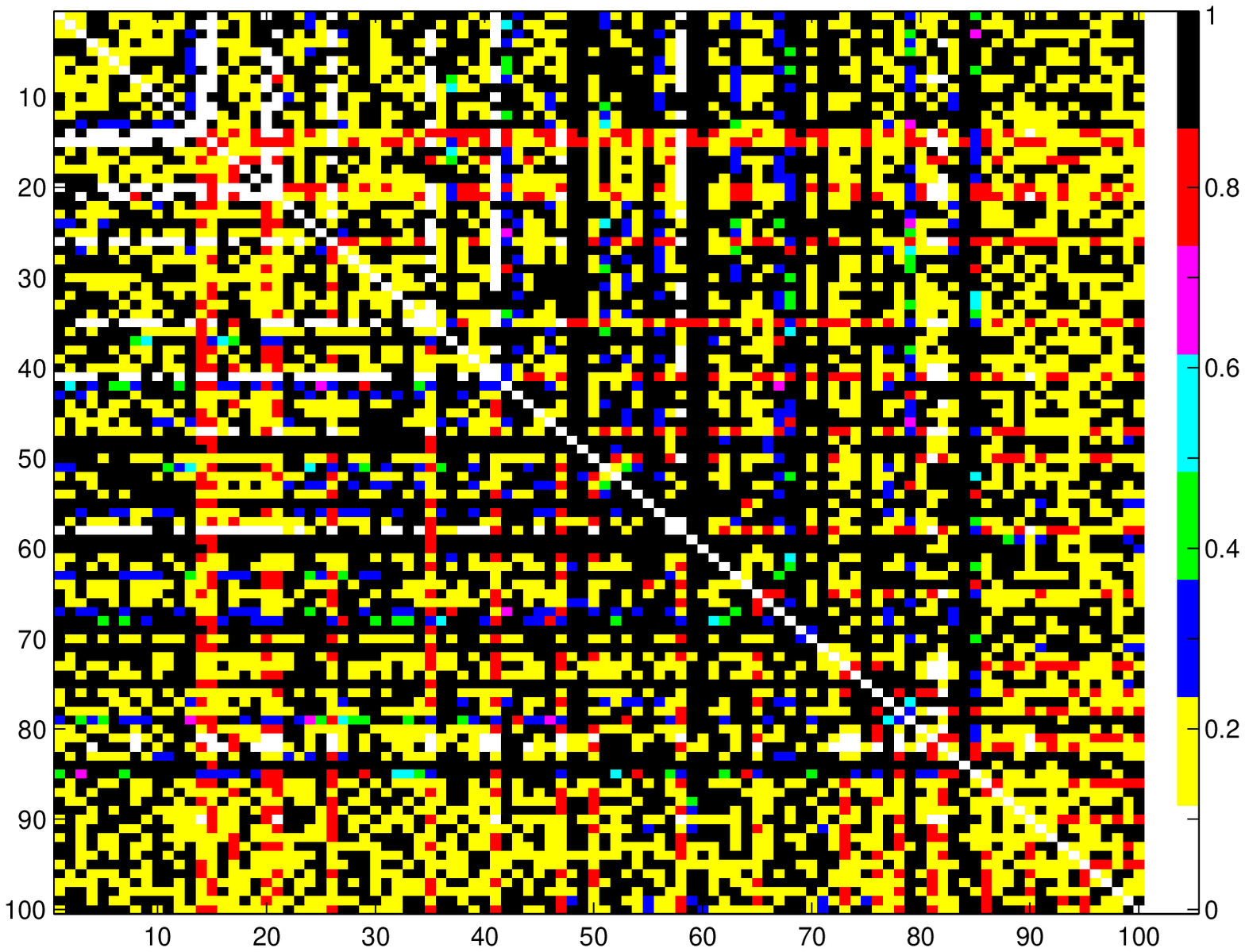,width=9truecm}
\end{center}
\caption{Same as in Fig.6 with $\alpha =1$ and $\beta =0.003$}
\end{figure}

\begin{figure}[htb]
\begin{center}
\psfig{figure=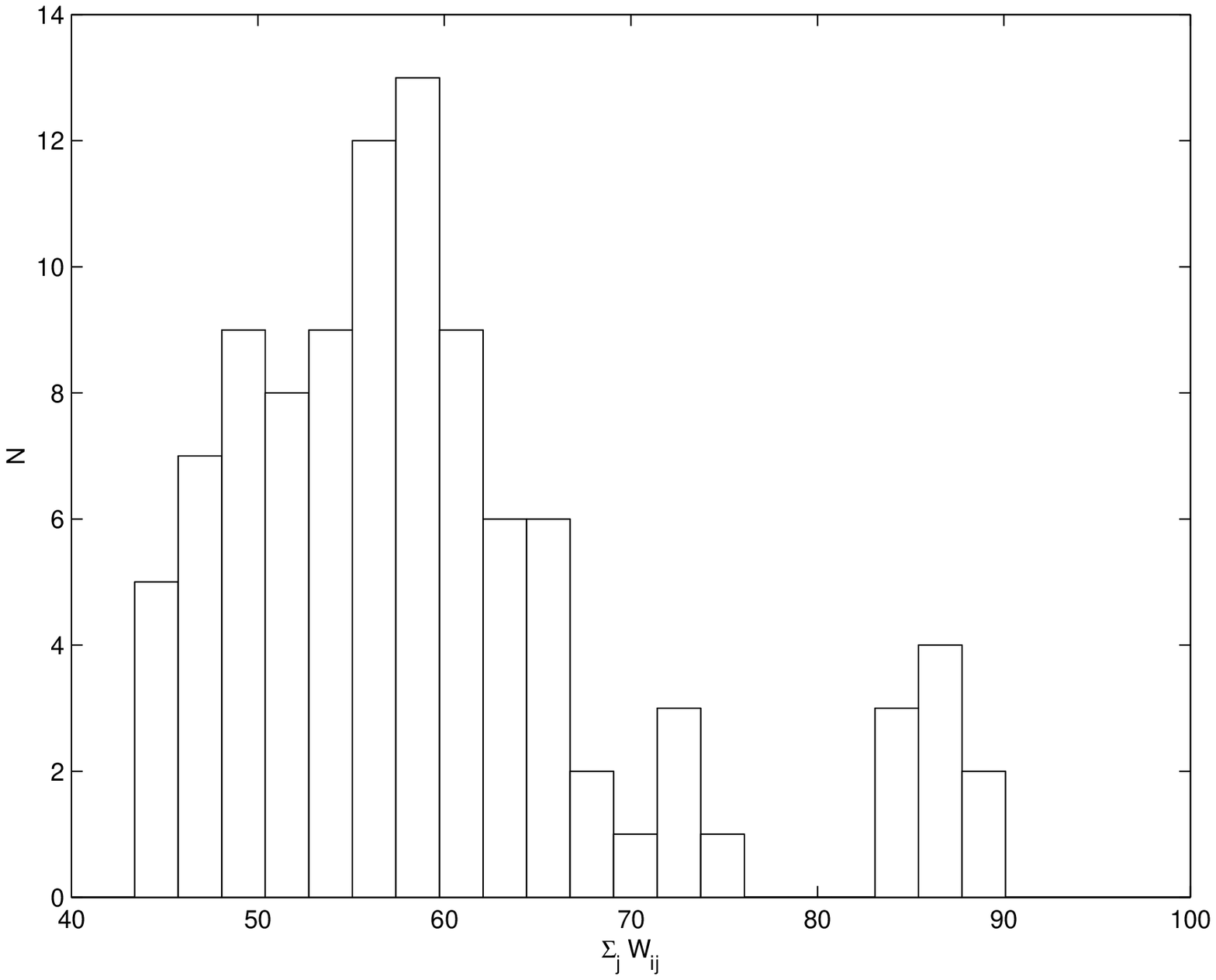,width=9truecm}
\end{center}
\caption{Degree distribution of the network in Fig.8}
\end{figure}

\begin{figure}[hbt]
\begin{center}
\psfig{figure=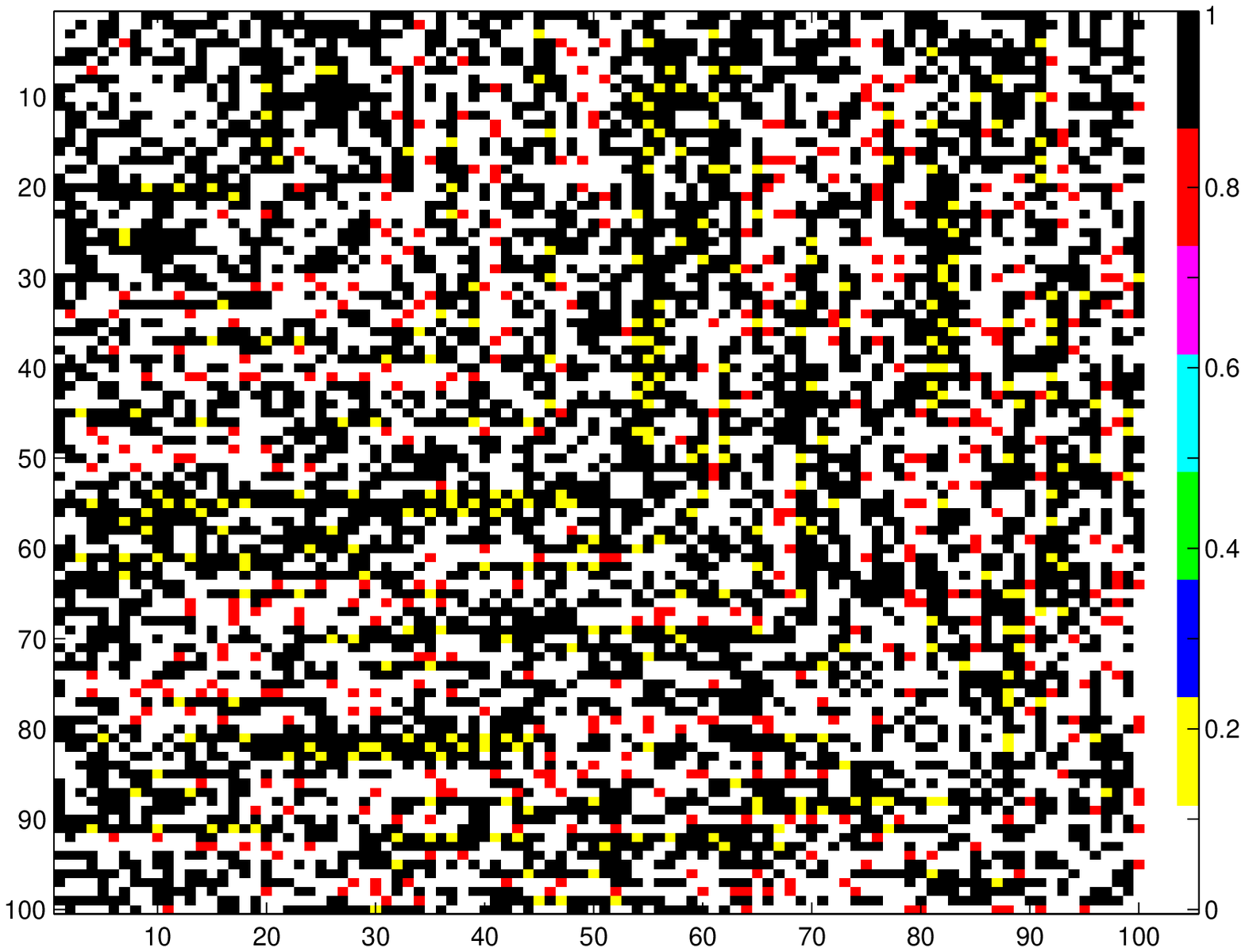,width=9truecm}
\end{center}
\caption{Typical equilibrium configuration of network connections evolved by
the potential $V_{2}\left( \left\{ W\right\} \right) $ ($\alpha =1,\beta
=0.05$)}
\end{figure}

Network evolution occurs either by the addition or elimination of
interactions between existing nodes or by the addition of new nodes. In both
cases, network evolution may be looked at as a dynamical system in the space
of network connections. In the case of growing networks, this dynamical
point of view may also be used by considering the evolution from zero of
previously vanishing connections.

This dynamical approach will be explored here. Using the global function
description, discussed in Section 2.1, two types of evolving networks will
be considered. The simplest situation occurs when the dynamics of the
connections is derived from a potential. In this case, exact expressions for
mean values and invariant measures may be obtained.

Consider 
\begin{equation}
V_{1}\left( \left\{ W\right\} \right) =\alpha \sum_{i<j}W_{ij}^{2}\left(
W_{ij}-1\right) ^{2}+\beta \sum_{i\neq j\neq l}\left( W_{ij}-1\right)
^{2}W_{jl}^{2}  \label{1.65}
\end{equation}
with the network evolving according to 
\begin{equation}
\frac{dW_{ij}}{dt}=-\frac{\partial V_{1}}{\partial W_{ij}}  \label{1.66}
\end{equation}
When $\alpha \neq 0$ and $\beta =0$, the connections evolve either to zero
or to one, depending on the initial conditions. Therefore the network (with $%
N$ nodes), as a dynamical system, is a multistable system with $2^{N\left(
N-1\right) /2}$ different equilibrium points. A typical configuration,
obtained from random initial conditions, is shown in Fig.6 ($N=100$) to
which corresponds the degree distribution shown in Fig.7.

When $\beta \neq 0$ the behavior is quite different, as shown in the typical
configuration of Fig.8 and degree distribution Fig.9. The degree $K_{i}$ of
a node $i$ is defined to be 
\begin{equation}
K_{i}=\sum_{j}W_{ij}  \label{1.68}
\end{equation}
holding for all intermediate values of $W_{ij}$.

One sees that for $\beta \neq 0$ some nodes are more connected than others. $%
V_{1}\left( \left\{ W\right\} \right) $ with $\beta \neq 0$ is a model for
preferential attachment.

It is not be practical to obtain mean values and distributions directly from
simulations. This being a multistable system many different simulations with
well distributed initial conditions would be required to obtain accurate
values. However, in this case, exact expressions may be obtained from the
unique invariant measure for the system with small random perturbations, as
discussed in Section 3.3 
\begin{equation}
\rho ^{\varepsilon }\sim \exp \left( -2\varepsilon ^{-2}V_{1}\left( \left\{
W\right\} \right) \right)  \label{1.69}
\end{equation}

As a second example consider 
\begin{equation}
V_{2}\left( \left\{ W\right\} \right) =\alpha \sum_{i<j}W_{ij}^{2}\left(
W_{ij}-1\right) ^{2}+\beta \sum_{i<j}\sum_{k\neq i,j}\frac{1}{\left|
i-j\right| }\left( W_{ik}^{2}+W_{jk}^{2}\right) \left( \left(
W_{ik}-1\right) ^{2}+\left( W_{jk}-1\right) ^{2}\right)  \label{1.70}
\end{equation}
For $\beta \neq 0$ a typical configuration is shown in Fig.10. The main
feature is the correlation between node connections. For $\alpha =1$, $\beta
=0.05$ and $N=100$ the sum of correlations between the node connections is
around $20$ whereas for $\alpha =1$, $\beta =0$ it is $\approx 4.5$. In
conclusion, $V_{2}\left( \left\{ W\right\} \right) $ is a model for
(approximate) node replication.

\section{Ergodic tools}

Topological and differential notions provide useful characterizations of the
overall structure of phase space. However, what is more important for the
applications is the dynamics in the phase space regions most frequently
visited by the system. This is provided by the ergodic theory, in particular
by the classification of invariant measures and their characterization by 
\textit{ergodic parameters.}

Let a dynamical system evolve on the support of a measure $\mu $ which is
left invariant by the dynamics. An \textit{ergodic parameter }$I_{F}\left(
\mu \right) $, characterizing the measure, is obtained whenever the
following limit 
\begin{equation}
I_{F}\left( \mu \right) =\lim_{T\rightarrow \infty }\frac{1}{T}%
\sum_{n=1}^{T}\digamma \left( f^{n}x_{0}\right)  \label{3.1}
\end{equation}
exists for $\mu -$almost every $x_{0}$. For continuous-time dynamics $f$
denotes the time-one map.

\subsection{Lyapunov and conditional exponents}

\textit{Lyapunov exponents} are the most widely used ergodic parameters.
More recently \textit{conditional exponents} have also been proposed as an
useful characterization of the dynamics.

Let $f:M\rightarrow M$\ , with $M\subset R^{m}$, $\mu $ a measure invariant
under $f$ and $\Sigma $ a splitting of $M$ induced by $R^{k}\times R^{m-k}$.
The \textit{conditional exponents}\ are the eigenvalues $\xi _{i}^{(k)}$ and 
$\xi _{i}^{(m-k)}$ of the limits 
\begin{eqnarray}
&&\lim_{n\rightarrow \infty }\left( D_{k}f^{n*}(x)D_{k}f^{n}(x)\right) ^{%
\frac{1}{2n}}  \label{3.2} \\
&&\lim_{n\rightarrow \infty }\left( D_{m-k}f^{n*}(x)D_{k}f^{n}(x)\right) ^{%
\frac{1}{2n}}  \nonumber
\end{eqnarray}
where $D_{k}f^{n}$\ and $D_{m-k}f^{n}$ are the $k\times k$\ and $m-k\times
m-k$ diagonal blocks of the full Jacobian. For $k=m$ , $\xi
_{i}^{(m)}=\lambda _{i}$ are the \textit{Lyapunov exponents}.

Proposed by Pecora \& Carroll [1990, 1991] to characterize synchronization
in chaotic systems, rigorous conditions for the existence of these limits
have been proven in [Vilela Mendes, 1998]. Existence $\mu $-almost
everywhere of both Lyapunov and conditional exponents is guaranteed by the
conditions of Oseledec's multiplicative ergodic theorem, in particular the
integrability condition, 
\begin{equation}
\int \mu (dx)\log ^{+}\left\| T(x)\right\| <\infty   \label{3.3}
\end{equation}
$T$ being either the Jacobian or its $k\times k$\ and $m-k\times m-k$
diagonal blocks. The set of points where the limit is defined has full
measure and 
\begin{equation}
\lim_{n\rightarrow \infty }\frac{1}{n}\log \left\| D_{k}f^{n}(x)u\right\|
=\xi _{i}^{(k)}  \label{3.4}
\end{equation}
\ with $0\neq u\in E_{x}^{i}/E_{x}^{i+1}$\ , $E_{x}^{i}$\ being the subspace
of $R^{k}$\ spanned by eigenstates corresponding to eigenvalues $\leq \exp
(\xi _{i}^{(k)})$.

Based on the spectra of Lyapunov and conditional exponents, several global
quantities have been defined to characterize self-organization and creation
of structures in networks of multiagent systems with arbitrary connection
structures. I list here the definitions and refer to [Vilela Mendes, 2000b,
2001] for proofs and examples.

\subsubsection{Structure index related to the Lyapunov spectrum}

A structure (in a collective system) is a phenomenon with a characteristic
scale very different from the scale of the elementary units in the system.
In a multi-agent system, a structure in space is a feature at a length scale
larger than the characteristic size of the agents and a structure in time is
a phenomenon with a time scale larger than the cycle time of the individual
agent dynamics. A (temporal)\textit{\ structure index} may then be defined by

\begin{equation}
S=\frac{1}{N}\sum_{i=1}^{N_{s}}\frac{T_{i}-T}{T}  \label{3.6}
\end{equation}
where $N$ is the total number of components (agents) in the coupled system, $%
N_{s}$ is the number of structures, $T_{i}$ is the characteristic time of
the structure $i$ and $T$ is the cycle time of the isolated agents (or,
alternatively the characteristic time of the fastest structure). A similar
definition applies for a \textit{spatial structure index}, by replacing
characteristic times by characteristic lengths.

Structures are collective motions of the system. Therefore their
characteristic times are the characteristic times of the separation
dynamics, that is, the inverse of the positive Lyapunov exponents. Hence,
for the \textit{temporal structure index}, one may write 
\begin{equation}
S=\frac{1}{N}\sum_{i=1}^{N_{+}}\left( \frac{\lambda _{0}}{\lambda _{i}}%
-1\right)  \label{3.7}
\end{equation}
the sum being over the positive Lyapunov exponents $\lambda _{i}$. $\lambda
_{0}$ is the largest Lyapunov exponent of an isolated component or some
other reference value.

The temporal structure index diverges whenever a Lyapunov exponent
approaches zero from above. Therefore the structure index diverges at the
points where long time correlations develop. Also, when in a multiagent
network the coupling between the agents increases, the positive part of the
Lyapunov spectrum contracts leading to an effective dimension reduction and
to partial synchronization effects [Vilela Mendes, 1999].

\subsubsection{Exponent entropies and dynamical selforganization}

Self-organization in a system concerns the dynamical relation of the whole
to its parts. The conditional Lyapunov exponents, being quantities that
separate the intrinsic dynamics of each component from the influence of the
other parts in the system, provide a \textit{measure of dynamical
selforganization} $I_{\Sigma }(\mu )$%
\begin{equation}
I_{\Sigma }(\mu )=\sum_{k=1}^{N}\left\{ h_{k}(\mu )+h_{m-k}(\mu )-h(\mu
)\right\}  \label{3.8}
\end{equation}
\ the sum being over all relevant partitions $\Sigma _{k}=R^{k}\times
R^{m-k} $ and$\,$

\[
h_{k}(\mu )=\sum_{\xi _{i}^{(k)}>0}\xi _{i}^{(k)};h_{m-k}(\mu )=\sum_{\xi
_{i}^{(m-k)}>0}\xi _{i}^{(m-k)};h(\mu )=\sum_{\lambda _{i}>0}\lambda _{i} 
\]
are the \textit{exponent entropies}, that is, the sums over positive
conditional and Lyapunov exponents.

$I_{\Sigma }(\mu )$ may also be given the following dynamical
interpretation: Lyapunov exponents measure the rate of information
production or, equivalently, they define the dynamical freedom of the
system, in the sense that they control the amount of change that is needed
today to have an effect on the future. In this sense the larger a Lyapunov
exponent is, the freer the system is in that particular direction, because a
very small change in the present state will induce a large change in the
future. The conditional exponents have a similar interpretation concerning
the dynamics as seen from the point of view of each agent and his
neighborhood [Vilela Mendes, 2000b]. However the actual information
production rate is given by the sum of the positive Lyapunov exponents, not
by the sum of the conditional exponents. Therefore, $I_{\Sigma }(\mu )$ is a
measure of apparent dynamical freedom (or apparent rate of information
production).

Being constructed as functions of well defined ergodic limits, both $%
I_{\Sigma }(\mu )$ and $S$ are also well defined ergodic parameters. They
characterize the dynamics of multiagent networks and, in addition, also
provide some insight on the relation between dynamics and the topology of
the network [Ara\'{u}jo \textit{et al.}, 2003].

\subsection{The problem of pattern discovery: Computational mechanics}

The ultimate practical goal in the study of dynamical systems is the
construction of models by which these systems might be predicted and (or)
controlled to some useful purpose. In extended systems with many degrees of
freedom, unless exact solutions are known, even a knowledge of the
(microscopic) equations of motion might not be very useful to predict the
collective features and patterns that the system generates. Crutchfield and
collaborators [Crutchfield \& Young, 1989] [Crutchfield, 1994] [Shalizi \&
Crutchfield, 2001] have developed a program of pattern discovery and
construction of minimal models inferred directly from the data generated by
the dynamical systems. Central to this approach is the notion of \textit{%
causal state}. Given the knowledge of the infinite past of a system, a
causal state is an equivalence class of pasts that have the same conditional
distribution of futures. Denoting by $\overleftarrow{s}$ and $%
\overrightarrow{s}$ the semi-infinite past and future time sequences of
coded states of the system, two past sequences $\overleftarrow{s_{1}}$ and $%
\overleftarrow{s_{2}}$ belong to the same \textit{causal state} if 
\[
P\left( \overrightarrow{s}|\overleftarrow{s_{2}}\right) =P\left( 
\overrightarrow{s}|\overleftarrow{s_{2}}\right) 
\]
for all $\overrightarrow{s}$(except perhaps in a zero measure set). The
dynamics of the system is then characterized by the set of causal states and
the transition probabilities between them. That is, the system is mapped
into a non-deterministic automaton called an $\varepsilon -$machine.
Minimality and uniqueness of the $\varepsilon -$machines has been proved.
Although more general, this approach bears some relation to the
reconstruction of hidden Markov processes and to grammatical inference.

As a tool for network dynamics this approach might be useful whenever
analytical equations are intractable or unknown. Reconstruction algorithms
for $\varepsilon -$machines were developed in some cases [Hanson \&
Crutchfield, 1997] [Crutchfield \& Feldman, 1997]. For extended system with
many degrees of freedom and irregular connections, one problem might be the
large size of the causal state alphabet. Nevertheless this is a very
interesting general approach that might be useful to map network dynamics
onto probabilistically equivalent automata.

\subsection{Construction of invariant measures}

In general a deterministic system has a multitude of invariant measures.
However, some of them have little practical interest, because they are not
stable for small random perturbations. Because systems in Nature are
subjected to perturbations, only the stable measures are \textit{physical
measures}. In some cases it is possible to use the properties of the
deterministic system to identify the physical measures. For example, in
Axiom A systems a unique physical measure may be identified with the
Sinai-Bowen-Ruelle (SBR)\ measure, a measure absolutely continuous along
unstable manifolds. However in most cases, for example in the multistable
systems so frequent in natural networks, the SBR characterization is
useless. Instead, it is better to study the stochastic differential equation
that is obtained from (\ref{1.1a}) by addition of a small noise term 
\begin{equation}
dx_{i}=X_{i}\left( x\right) dt+\varepsilon \sigma \left( X\right) dW_{t}
\label{3.9}
\end{equation}
$W_{t}$ being a Wiener process and $\sigma \left( X\right) $ a $X-$dependent
diffusion coefficient. A great deal of information on the invariant measure
for this process may be obtained using the theory of small random
perturbations of dynamical systems [Freidlin \& Wentzell, 1984, 1994]
[Kifer, 1974].

If, in the decomposition (\ref{1.3}), $X\left( x\right) $ has only a
gradient component, an explicit form for the invariant measure may be
obtained. If 
\begin{equation}
X\left( x\right) =-\nabla _{(g)}V\left( x\right)  \label{3.10}
\end{equation}
$\nabla _{(g)}$ being the gradient in the metric 
\begin{equation}
ds^{2}=\sum a_{ij}\left( x\right) dx_{i}dx_{j}  \label{3.11}
\end{equation}
with $\sigma \left( x\right) $ in (\ref{3.9}) chosen such that 
\begin{equation}
a_{ij}\left( x\right) =\left( \sigma \left( x\right) \sigma ^{*}\left(
x\right) \right) _{ij}^{-1}=g_{ij}\left( x\right)  \label{3.12}
\end{equation}
then, the density of the invariant measure is 
\begin{equation}
\rho ^{\varepsilon }\left( x\right) =C_{\varepsilon }\exp \left(
-2\varepsilon ^{-2}V\left( x\right) \right)  \label{3.13}
\end{equation}
as may be easily checked from the forward Kolmogorov equation. In this case,
finding the stable minima and level sets of $V\left( x\right) $ one
characterizes the multistability of the network, their basins of attraction
and, from the values of $V\left( x\right) $ in these sets, the relative
occurrence probability of each attractor.

For general $X\left( x\right) $, small $\varepsilon $ estimates of the
invariant measure for (\ref{3.9}) are also possible. Here the crucial role
is played by the functional 
\begin{equation}
S_{oT}\left( \varphi \right) =\frac{1}{2}\int_{0}^{T}\sum_{ij}a_{ij}\left(
\varphi _{t}\right) \left( \stackrel{\bullet }{\varphi _{t}^{i}}-X^{i}\left(
\varphi _{t}\right) \right) \left( \stackrel{\bullet }{\varphi _{t}^{j}}%
-X^{j}\left( \varphi _{t}\right) \right) dt  \label{3.14}
\end{equation}
and the infimum 
\begin{equation}
U\left( x,y\right) =\inf \left\{ S_{0T}\left( \varphi \right) :\varphi
_{0}=x,\varphi _{T}=y,t\in [0,T]\right\}  \label{3.15}
\end{equation}
taken over intervals $[0,T]$ of arbitrary length.

An equivalence relation is established between points in the domain by $%
x\sim y$ if $U\left( x,y\right) =U\left( y,x\right) =0$. Let the domain be
partitioned into a number of compacta $\left\{ K_{i}\right\} $ with each $%
\omega -$limit set of the deterministic dynamics contained entirely in one
compactum and $x\sim y$ inside each compactum. Then, the (small $\varepsilon 
$) asymptotics of the invariant measure is obtained from the invariant
measure of the Markov chain of transitions between the compacta. For
sufficiently small $\varepsilon $ the measure of each compactum is
approximated by 
\begin{equation}
\exp \left\{ -\varepsilon ^{-2}\left( W\left( K_{i}\right) -\min_{i}W\left(
K_{i}\right) \right) \right\}   \label{3.16}
\end{equation}
where 
\begin{equation}
W\left( K_{i}\right) =\min_{g\in G\left( i\right) }\sum_{\left( m\rightarrow
n\right) \in g}V\left( K_{m},K_{n}\right)   \label{3.17}
\end{equation}
$V\left( K_{m},K_{n}\right) $ is the minimum of the function (\ref{3.15})
between points in compacta $K_{m}$ and $K_{n}$ and the sum runs over graphs
that have exactly one closed cycle and this cycle contains the compactum $%
K_{i}$. For proofs I refer to [Freidlin \& Wentzell, 1984].

\subsection{A family of ergodic parameters}

Ergodic parameters like the Lyapunov and the conditional exponents, are
global functions of the invariant measure. However, the invariant measure
itself contains more information. Ergodic parameters being defined by
infinite-time limits, these quantities will fluctuate and, in general,
fluctuations will not be Gaussian. The quantity describing the fluctuations
is again an ergodic parameter and the same reasoning applies in turn to its
fluctuations, etc. [Ruelle, 1987]. Therefore, to characterize the measure, a
larger set of parameters is needed. To construct this larger set from the
fluctuations is not very practical and a different approach will be followed
here, namely a variational approach.

In a restricted sense, a variational principle states that the equations of
motion may be written in the form $\delta S=0$, where $S$ is a functional of
the dynamical variables and $\delta $ is the Gateaux derivative. Only a
limited set of dynamical systems may be described by a variational principle
in this restricted sense. However, if one only requires that $\delta S=0$
and the equations of motion possess the same set of solutions, essentially
all differential equation problems admit a variational formulation [Tonti,
1992]. Let 
\begin{equation}
\stackrel{\bullet }{x}_{i}=X_{i}\left( x\right)   \label{3.19}
\end{equation}
be a differentiable continuous-time dynamical system and $S$ be the
functional 
\begin{equation}
S=\int \int_{0}^{T}dtd\tau \sum_{i}\left\{ \stackrel{\bullet }{x}_{i}\left(
t\right) -X_{i}\left( x\left( t\right) \right) \right\} g\left( t,\tau
\right) \left\{ \stackrel{\bullet }{x}_{i}\left( \tau \right) -X_{i}\left(
x\left( \tau \right) \right) \right\}   \label{3.20}
\end{equation}
where $g\left( t,\tau \right) $ is a symmetric kernel ($g\left( t,\tau
\right) =g\left( \tau ,t\right) $). Let us compute the Gateaux derivative
for variations in space restricted by the boundary conditions 
\begin{equation}
u\left( 0\right) =u\left( T\right) =0  \label{3.21}
\end{equation}
From 
\begin{equation}
\delta _{u}S=-\int \int_{0}^{T}dtd\tau \sum_{i,k}u_{k}(t)\left\{ \delta _{ki}%
\frac{dg\left( t,\tau \right) }{dt}-\partial _{k}X_{i}\left( x\left(
t\right) \right) g\left( t,\tau \right) \right\} \left\{ \stackrel{\bullet }{%
x}_{i}\left( \tau \right) -X_{i}\left( x\left( \tau \right) \right) \right\} 
\label{3.22}
\end{equation}
we have

\textit{Lemma}: The equations of motion (\ref{3.19}) and the critical points
of the functional ($\delta _{u}S=0$) have the same set of solutions if 
\begin{equation}
K\left( t,\tau \right) =\delta _{ki}\frac{dg\left( t,\tau \right) }{dt}%
-\partial _{k}X_{i}\left( x\left( t\right) \right) g\left( t,\tau \right)
\label{3.23}
\end{equation}
is invertible.

\textit{Remarks}:

a) If $K\left( t,\tau \right) $ is not invertible, the solutions of the
equations of motion are still critical points of the functional, but this
one might have other solutions.

b) A variational principle, with only $u\left( 0\right) =0$ being required,
may also be obtained by choosing a kernel such that $g\left( t,T\right) =0$.

The critical points of the $S$ functional contain the same information as
the equations of motion. Therefore the dynamics may be characterized by the
properties of the critical points, in particular by their Hessian matrix.
Computing the second Gateaux derivative on the orbits one obtains 
\begin{equation}
\left. \delta _{u,v}^{2}S\right| _{\delta S=0}=\int \int_{0}^{T}dtd\tau
\sum_{i,j}u_{i}\left( t\right) v_{j}\left( \tau \right) H_{ij}\left( t,\tau
\right)  \label{3.24}
\end{equation}
with 
\begin{equation}
H_{ij}\left( t,\tau \right) =\sum_{k}\left\{ 
\begin{array}{c}
\frac{dg\left( t,\tau \right) }{dt}\partial _{j}X_{i}\left( x\left( \tau
\right) \right) -\partial _{i}X_{k}\left( x\left( t\right) \right) g\left(
t,\tau \right) \partial _{j}X_{k}\left( x\left( \tau \right) \right) \\ 
+\frac{d^{2}g\left( t,\tau \right) }{dtd\tau }\delta _{ij}+\partial
_{i}X_{j}\left( x\left( t\right) \right) \frac{dg\left( t,\tau \right) }{%
d\tau }
\end{array}
\right\}  \label{3.26}
\end{equation}
Now assume that the symmetric kernel $g\left( t,\tau \right) $ is a function
of finite support of $t-\tau $%
\begin{equation}
g\left( t,\tau \right) =g\left( t-\tau \right) =0\qquad \mathnormal{%
for\qquad }\left| t-\tau \right| >r  \label{3.27}
\end{equation}
Define 
\begin{equation}
J_{0,T}^{(n)}=\int \int_{0}^{T}\mathnormal{Tr}\left( H^{n}\left( t,\tau
\right) \right) dtd\tau  \label{3.27a}
\end{equation}
as well as 
\[
J_{0,T}^{(n)\prime }=J_{0,T}^{(n)}+\int \int_{T-r}^{T+r}\left| \mathnormal{Tr%
}\left( H^{n}\left( t,\tau \right) \right) \right| dtd\tau +\int
\int_{-r}^{r}\left| \mathnormal{Tr}\left( H^{n}\left( t,\tau \right) \right)
\right| dtd\tau 
\]
Then 
\[
J_{0,T_{1}+T_{2}}^{(n)\prime }\leq J_{0,T_{1}}^{(n)\prime
}+J_{T_{1},T_{2}}^{(n)\prime } 
\]
and we are in the conditions of Kingman's sub-additive ergodic theorem.
Taking limits, if both $X_{i}$ and $\partial _{i}X_{k}$ are bounded $%
J_{0,T}^{(n)\prime }$and $J_{0,T}^{(n)}$ differ only by a finite quantity
and one concludes:

\textit{Theorem}: If $\mu $ is an invariant measure of the dynamics in (\ref
{3.19}), $X_{i}$ and $\partial _{i}X_{k}$ are bounded and there is $M\geq 0$
such that $J_{0,T}^{(n)}\geq -M$ for sufficiently large $T$, then the limit 
\begin{equation}
I_{n}\left( \mu \right) =\lim_{T\rightarrow \infty }\frac{1}{T}J_{0,T}^{(n)}
\label{3.27b}
\end{equation}
exists and 
\[
\int \lim_{T\rightarrow \infty }\frac{1}{T}J_{0,T}^{(n)}d\mu
=\lim_{T\rightarrow \infty }\frac{1}{T}\int J_{0,T}^{(n)}d\mu 
\]

$I_{n}\left( \mu \right) $ for $n=1,2,\cdots $ is a family of ergodic
parameters for the $\mu -$measure preserving dynamics.

A similar construction for discrete-time maps may be found in [Vilela
Mendes, 1984] [Carreira \textit{et al.}, 1991].

\subsection{Synchronization, mode-locking and dynamical correlations}

The onset of correlated motions in coupled many-agent systems is a
phenomenon of widespread occurrence in many scientific fields. The most
dramatic effect is the synchronization of assemblies of coupled dynamical
systems which, when in isolation, may have quite different rhythms [Pikovski 
\textit{et al.}, 2001]. Examples are biological rhythms [Winfree, 1967] like
the pacemaker cells in the heart [Peskin, 1975], neural systems [Golomb \&
Hansel, 2000], synchronous metabolism [Aldridge \& Pye, 1976], flashing
fireflies [Buck, 1988], laser arrays [Jiang \& McCall, 1993], even fads and
social trends may be interpreted as synchronization of distinct agent
dynamics. The study of the correlated behavior of many-agent dynamics is
also closely related to the problem of control in extended dynamical systems.

I will consider both the coupled behavior of non-chaotic systems
(oscillators with distinct individual frequencies) and of systems with
isolated chaotic dynamics. In both cases one may distinguish between
globally coupled systems and systems where each agent has a limited range or
number of interacting partners.

\begin{figure}[htb]
\begin{center}
\psfig{figure=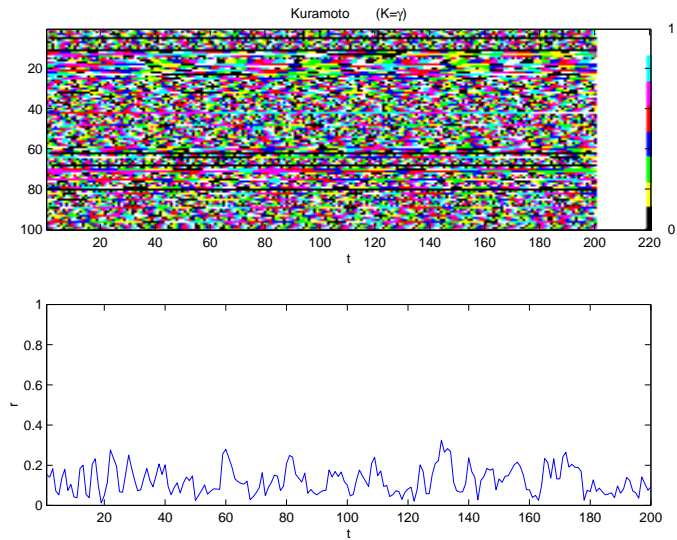,width=9truecm}
\end{center}
\caption{A Kuramoto system below threshold}
\end{figure}
\begin{figure}[hbt]
\begin{center}
\psfig{figure=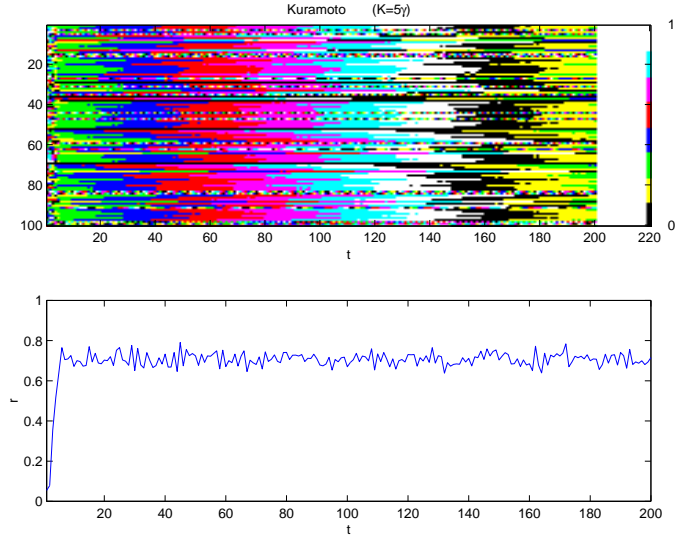,width=9truecm}
\end{center}
\caption{A Kuramoto system above threshold}
\end{figure}

For systems of oscillators the canonical example is the Kuramoto model
[Kuramoto, 1984, 1991], 
\begin{equation}
\frac{d\theta _{i}}{dt}=\omega _{i}+\frac{K}{N-1}\sum_{j=1}^{N}\sin \left(
\theta _{j}-\theta _{i}\right)   \label{S1}
\end{equation}
with $K>0$ and the frequencies $\omega _{i}$ randomly distributed around a
central value $\omega _{0}$ with the shifted Cauchy distribution 
\begin{equation}
p\left( \omega \right) =\frac{\gamma }{\pi \left[ \gamma ^{2}+\left( \omega
-\omega _{0}\right) ^{2}\right] }  \label{S3}
\end{equation}
A great deal of work has been done on this model (for a review see Strogatz
[2000]). The existence of a synchronized cluster is characterized by the
order parameter 
\begin{equation}
r\left( t\right) =\left| \frac{1}{N}\sum_{j=1}^{N}e^{i\theta _{j}\left(
t\right) }\right|   \label{S4}
\end{equation}

It is found that in the $N\rightarrow \infty $ and $t\rightarrow \infty $
limit, $r=0$ for $K<2\gamma $ and $r=\sqrt{1-\left( 2\gamma /K\right) }$ for 
$K\geq 2\gamma $. That is, there is a coupling threshold above which part of
the oscillators starts to synchronize. Figs.11 and 12 show the
nonsynchronized (at $K=\gamma $) and the synchronized (at $K=5\gamma $)
behavior for $100$ oscillators. The upper plot displays the color-coded
values of the oscillator variables at the end of each unit time interval.
The lower plots show the time evolution of the order parameter. Fig.13
compares the numerically computed Lyapunov spectrum in the synchronized and
non-synchronized situations. One sees that even below the synchronization
threshold ($K=2\gamma $), part of the Lyapunov exponents becomes negative,
meaning that there are many contracting directions, implying an effective
dimension-reduction of the asymptotic behavior of the system. This clearly
suggests that synchronization is not the whole story and that even before
synchronization strong correlations must develop between the dynamics of the
individual oscillators. 
\begin{figure}[htb]
\begin{center}
\psfig{figure=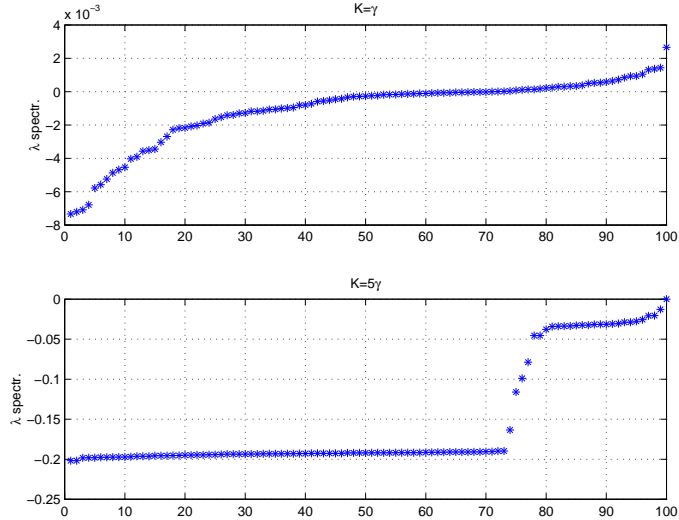,width=9truecm}
\end{center}
\caption{Lyapunov spectrum below and above threshold for the Kuramoto system}
\end{figure}
\begin{figure}[hbt]
\begin{center}
\psfig{figure=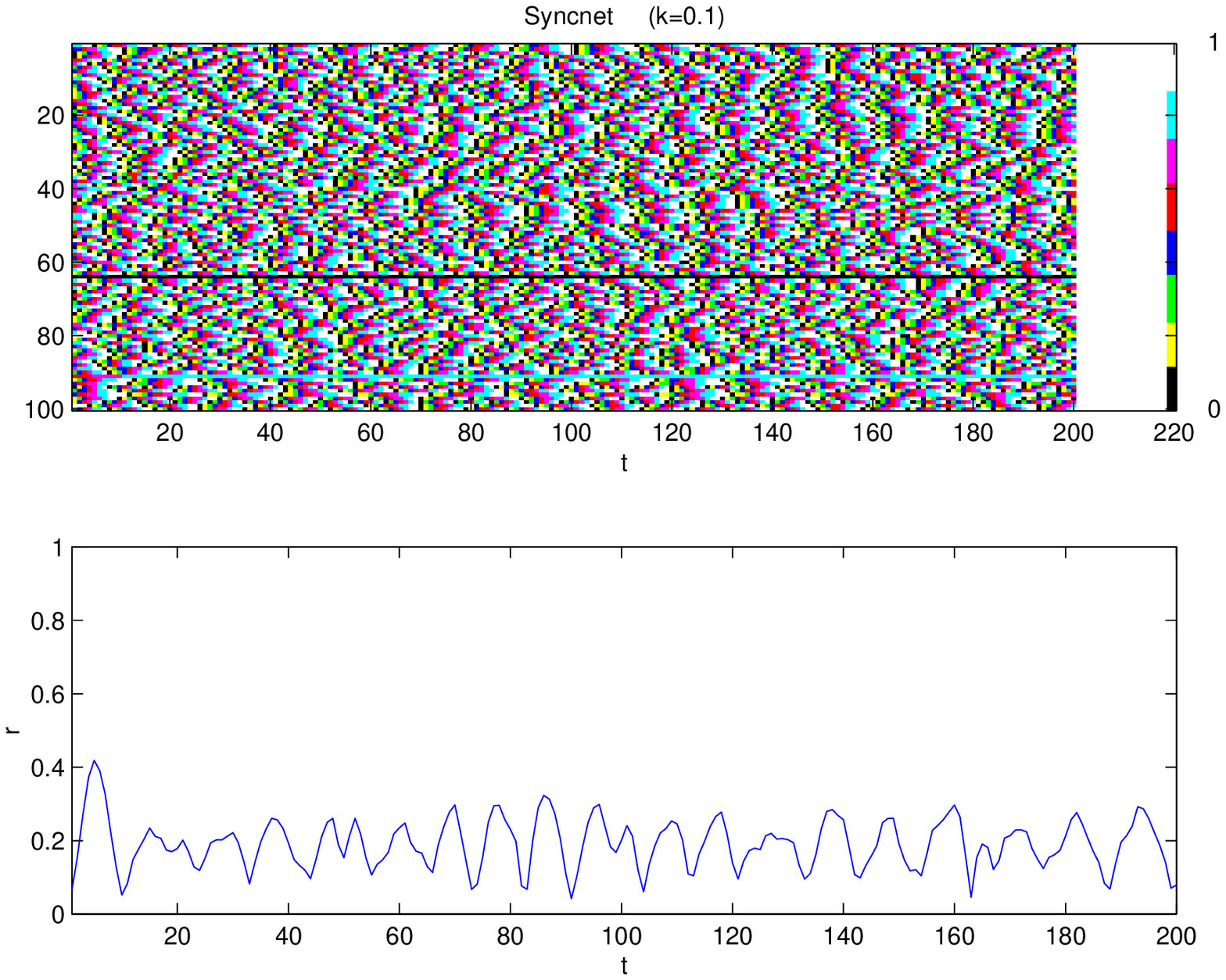,width=9truecm}
\end{center}
\caption{Non-synchronized behavior of the discrete-time oscillators (Eq.\ref
{S5})}
\end{figure}

\begin{figure}[htb]
\begin{center}
\psfig{figure=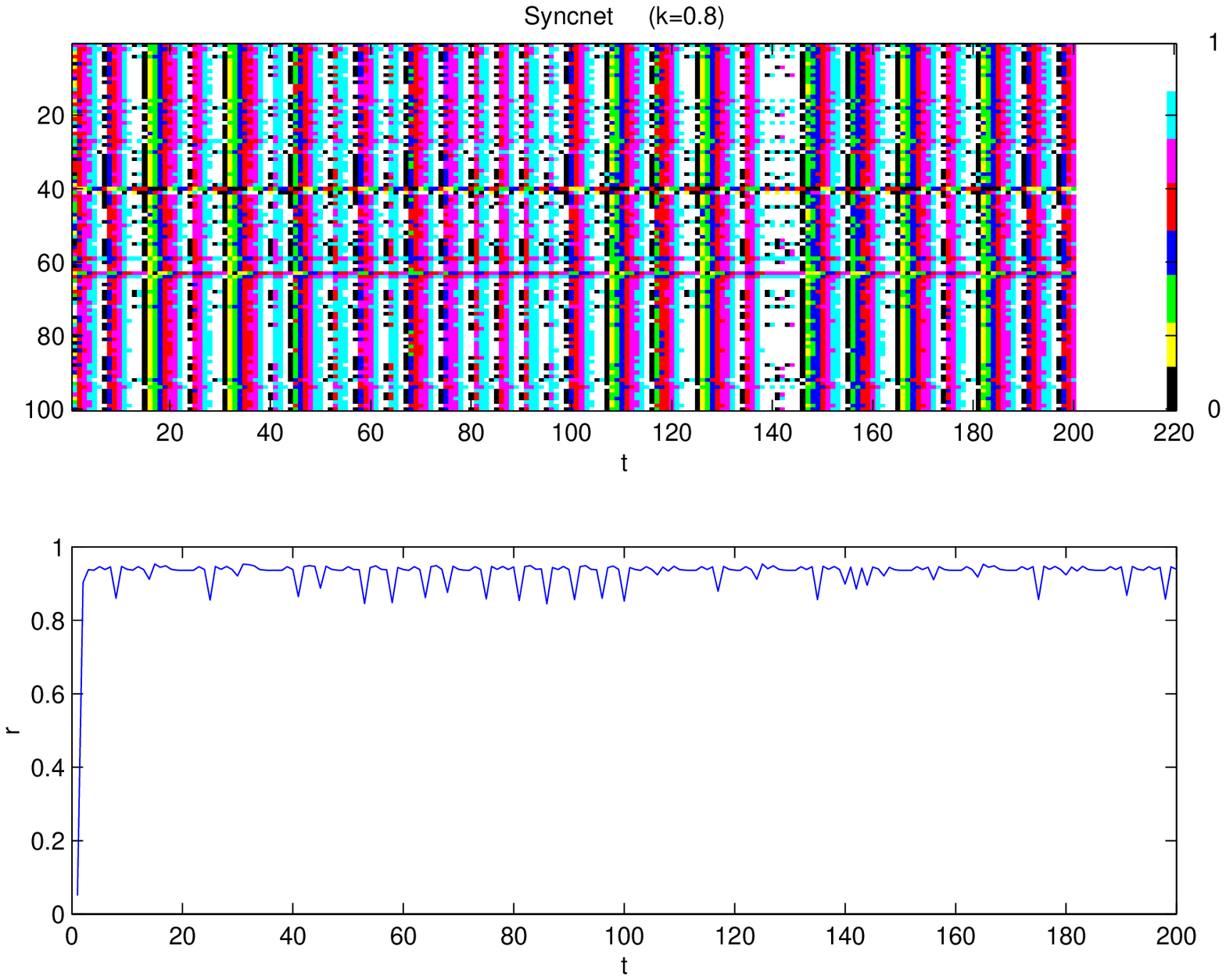,width=9truecm}
\end{center}
\caption{Synchronized behavior of the discrete-time oscillators (Eq.\ref{S5})
}
\end{figure}
\begin{figure}[hbt]
\begin{center}
\psfig{figure=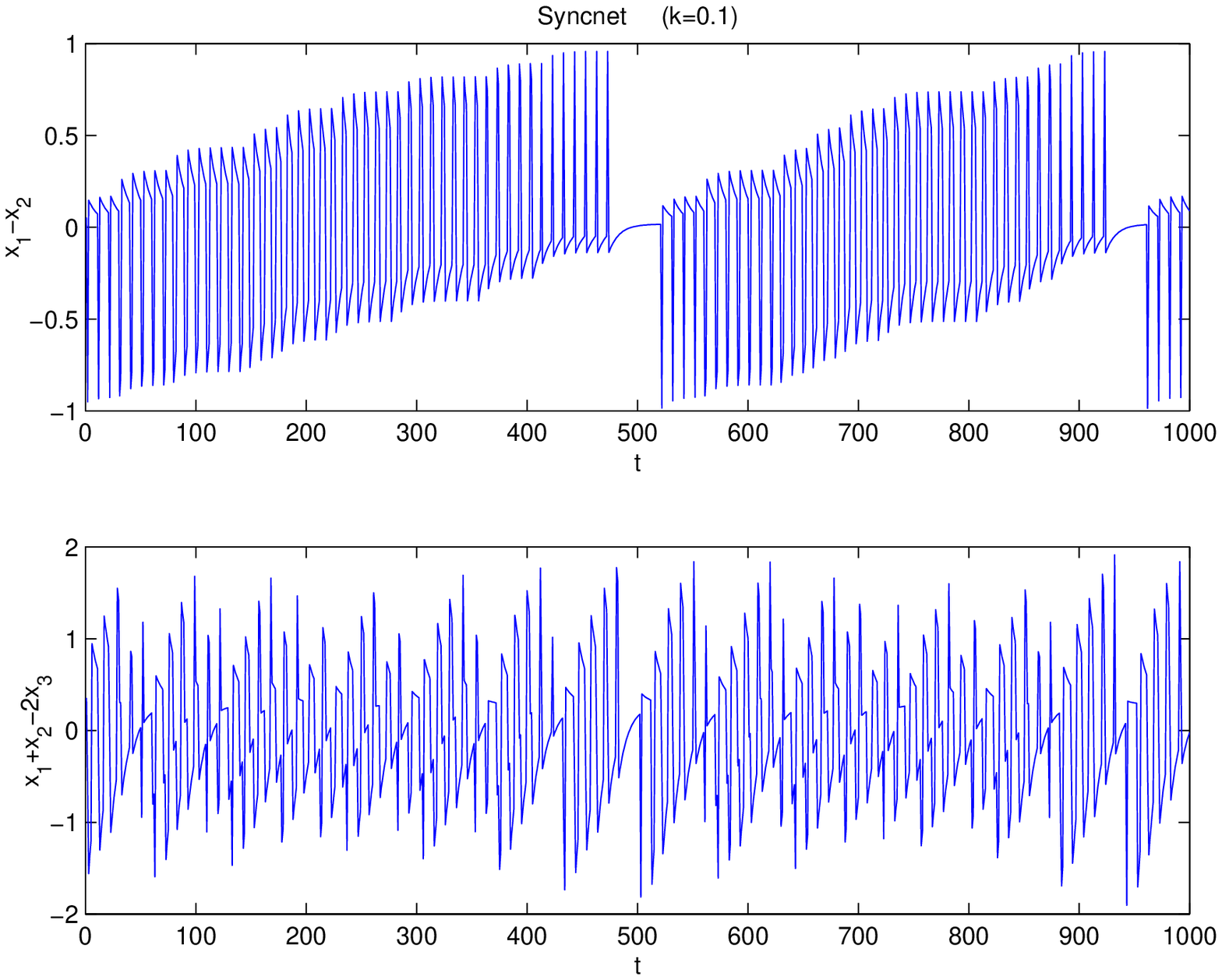,width=9truecm}
\end{center}
\caption{Correlations in the discrete-time oscillators system.}
\end{figure}

A type of correlation, of which synchronization is a limiting case is 
\textit{mode-locking}. Mode-locking is the entrainment of some integer
combination of the frequencies to zero. It also plays an important role in
the dynamics of coupled oscillators [MacKay, 1994]. However even if all the
effective frequencies are incommensurable, the existence of negative
Lyapunov directions, implies the existence of dynamical correlations between
the oscillators. What is important is the dimension of the invariant measure
and the correlations may be characterized by the eigenvectors of the
Lyapunov spectrum. These notions are better clarified in a simple model with
exactly computable Lyapunov spectrum. Let the dynamics of an assembly of
discrete-time oscillators be 
\begin{equation}
x_{i}\left( t+1\right) =x_{i}\left( t\right) +\omega _{i}+\frac{k}{N-1}%
\sum_{j=1}^{N}f_{\alpha }\left( x_{j}-x_{i}\right)   \label{S5}
\end{equation}
with $x_{i}\in [0,1)$ and $f_{\alpha }\left( x_{j}-x_{i}\right) =\alpha
\left( x_{j}-x_{i}\right) $ $\left( \mathnormal{mod}1\right) $ and the $%
\omega _{i}^{\prime }s$ distributed according to $p\left( \omega \right) $,
as above.

The Lyapunov spectrum is composed of one isolated zero and $\log \left( 1-%
\frac{N}{N-1}\alpha k\right) $ ($N-1$)-times. However, although (for all $%
k>0 $) $N-1$ contracting directions are always present, it is only for
sufficiently large $k$ that synchronization effects emerge as shown in
Figs.14 and 15. Nevertheless dynamical correlations do exist for all $k$, no
matter how small and the Lyapunov dimension is always one. In this case, the
eigenvectors of the Lyapunov spectrum may be exactly computed and the
correlations explicitly identified. This is illustrated in Fig.16.

So far I have dealt with coupled oscillators, that is, with systems which
have individual nonchaotic dynamics. Another important field with many
practical applications refers to the case where the individual node dynamics
is chaotic. Synchronization of chaotic systems has been extensively studied
(for a review see Pecora \textit{et al.} [1999]) and is still a field of
current research [Wei \textit{et al.}, 2002]. However, as in the
oscillators, for networks of chaotic elements the interesting phenomena go
beyond synchronization effects. Correlations and self-organization effects
may be characterized by ergodic parameters. I refer to [VilelaMendes, 1999,
2000b] for illustrative examples of networks of chaotic elements both
globally connected and with a limited range of interactions. The Lyapunov
spectrum and the entropies associated to the conditional exponents provide a
characterization of the emergent phenomena. It should be noticed that
dynamical correlations play an important role in the organization of the
dynamics even when there is no reduction of the Lyapunov dimension [Vilela
Mendes, 1999]. As before these correlations are associated to the
eigenvectors of the Lyapunov spectrum.

\subsection{Dynamics and network topology}

The topology of the network connections has a determining effects on the
dynamical phenomena taking place in the network [Watts, 1999]. Local
clustering and far-reaching connections in the network influence the spread
of infectious diseases [Boots \& Sasaki, 1999] [Keeling, 1999]
[Pastor-Satorras \& Vespignani, 2002] [Lloyd \& May, 2001] or social fads.
The nature and range of the couplings influences the speed of signal
propagation and even the computational abilities of the network. On the
other hand the topological structure of the networks by influencing the
dynamics, may have a feedback effect on network growing and therefore also
evolutionary significance.

In particular the small world topology [Watts \& Strogatz, 1998] [Watts,
1999] [Strogatz, 2001] (with both small path length and large clustering)
has been found to modify or enhance coherent behavior effects
[Lago-Fernandez \textit{et al.}, 2000] [Gade \& Hu, 2000] and in general
influence the dynamics in the network [Yang, 2001] [Kulkarni \textit{et al.}%
, 1999]. An attempt has also been made to relate the ergodic parameters of a
dynamical system, living on the network, with the changes of topological
structure associated to the small-world features. It turns out that whereas
the emergence of short path lengths is associated to a modification of the
Lyapunov spectrum, the transition from the small world to the random regimen
is characterized by the conditional exponent entropies (for details see Ara%
\'{u}jo \textit{et al.} [2003]).

\section{The logic approach to network dynamics}

Thomas and collaborators [Thomas \& D'Ari, 1990] [Thomas, 1991] [Thieffry 
\textit{et al.}, 1993] [Snoussi \& Thomas, 1993] [Thomas \textit{et al.},
1995] have developed logical tools to analyze biological regulatory
networks. This treatment seems particularly appropriate to deal with
regulatory networks where most interactions are characterized by a threshold
and a saturation plateau. The regulator is usually inefficient below a
threshold concentration and its effect rapidly levels off above the
threshold.

The elements of the network and their interactions are represented by
discrete \textit{variables}, \textit{functions} and \textit{parameters}.
Because some variables have several distinct actions, one often needs to
consider more than two logical levels. One may also consider the threshold
values $s^{(i)}$ separating the logical values. Then, the most general
logical variable takes values in the set 
\begin{equation}
\left\{ 0,s^{(1)},1,s^{(2)},2,\cdots \right\}  \label{4.1}
\end{equation}
The state of the system is described by the \textit{state} vector $xyz\cdots 
$ containing the logical values of the variables and the evolution of the
system by a vector of \textit{functions} $XYZ\cdots $ representing the
excitatory or inhibitory interactions in the network. For example, for the
graph of interactions 
\begin{equation}
\underset{\underset{-}{\leftarrow ------}}{\underset{|}{x}\stackrel{+}{%
\rightarrow }y\stackrel{-}{\rightarrow }\underset{|}{z}}  \label{4.2}
\end{equation}
the logical functions are 
\begin{equation}
X=\stackrel{-}{z},\qquad Y=x,\qquad Z=\stackrel{-}{y}  \label{4.3}
\end{equation}
$X=0$ means that the product $x$ is not being produced and $X=1$ means that
the product $x$ is being produced, not that $x=1$ immediately. For example,
if the initial state is $000$, after a certain time the state might become $%
100$ or $001$, depending on whether the time delay $\left( t_{x}\right) $
for production of $x$ is smaller or larger than the time delay $\left(
t_{z}\right) $ for the production of $z$. Notice that although the \textit{%
image} $XYZ$ of the state $000$ is $101$, the next state is either $100$ or $%
001$ because it is highly unlikely that $t_{x}=t_{z}$. The state $001$ is a 
\textit{stable state} because it equals its image, whereas $100$ further
evolves to $110$ or $101$, of which $110$ is stable but $101$ is not, etc.

A central role in this formalism is played by the oriented circuits (closed
series of interactions where each element appears only once), called \textit{%
feedback loops}. Feedback loops are positive or negative according to
whether they have an even or odd number of negative interactions. Positive
loops generate multistability and negative loops generate homeostasis, that
is, the variables in the loop tend to middle range values, with or without
oscillations.

The role of the \textit{parameters} in the logical approach is to allow for
the distinction between weak and strong interactions. Therefore the
parameters are actually real values, not logical variables. For example, in
the following interaction graph 
\begin{equation}
x\rightleftarrows _{-}^{+}y\sqsupset ^{+}  \label{4.4}
\end{equation}
$x$ has two values $\left( 0,1\right) $ because it has one action only and $%
y $ has three values $\left( 0,1,2\right) $ because it has two actions and
thus two thresholds. The corresponding logical functions are 
\begin{equation}
X=\stackrel{-}{y^{1}},\qquad Y=x^{1}+y^{2}  \label{4.5}
\end{equation}
in which all the variables are Boolean. The function $Y$ means that if $x$
is above its threshold or $y$ above the second threshold ($x^{1}=1$ or $%
y^{2}=1$) then $y$ is going to be produced. However we might give different
weights to the variables by writing 
\begin{equation}
X=d_{x}\left( K_{1}\stackrel{-}{y^{1}}\right) ,\qquad Y=d_{y}\left(
K_{2}x^{1}+K_{3}y^{2}\right)  \label{4.6}
\end{equation}
where the $K_{i}^{^{\prime }}s$ are real variables and $d_{x}$ and $d_{y}$
operators that discretize the value in brackets. Here the $+$ sign is the
real algebraic sum not the logical one.

States with variable values $xyz\cdots $ that involve only logical values
are called \textit{regular states} and those which involve some threshold
values are called \textit{singular states}. It is found that each feedback
loop can be characterized by a singular logical state located on the
thresholds. For appropriate parameter values, this state is stationary and
the corresponding loop \textit{functional}. In this context, functional
means that if the loop is positive it actually produces multistability and
if negative it generates homeostasis. A technique for the analysis of the
network consists in dissociating it into its feedback loops and checking the
dynamics of each loop.

Other logical approaches have been developed to analyze networks: the
Boolean network models [Kaufman, 1993] [Somogyi \& Sniegoski, 1996], hybrid
models using logical and continuous variables [Glass, 1975] [Lewis \& Glass,
1991] and rule-based formalisms [Brutlag \textit{et al.}, 1991] [Meyers \&
Friedland, 1984] [Shimada \textit{et al.}, 1995] [Hofest\"{a}dt \& Meineke,
1995].

A general problem in the logical approaches to network dynamics is to
establish the correspondence of the logical description to the corresponding
set of differential equations. For the multilevel approach (with parameters)
of Thomas and collaborators, if the thresholds are represented by Hill
functions, the correspondence becomes perfect when the Hill functions become
very steep. The problem of establishing the correspondence between logical
descriptions and smooth dynamical systems has also been addressed in other
contexts. For example, in [Martins \& Vilela Mendes, 2001] a correspondence
is established between a type of neural networks and a logical programming
language and in [Martins \textit{et al.}, 2001] a correspondence between
controlled dynamical systems and context-dependent languages. Similar
techniques may be used for general networks [Aguirre et al., 2004].

{\LARGE References}

Abarbanel, H. D. I. \& Rouhi, A. [1987] ``Hamiltonian structures for smooth
vector fields'', \textit{Phys. Lett. A }124, 281-286.

Afraimovich, V. \& Fernandez, B. [2000] ``Topological properties of linearly
coupled expanding map lattices'', \textit{Nonlinearity} 13, 973-993.

Aguirre, C., Martins, J. \& Vilela Mendes, R. [2004] ``From differential
equations to multilevel logic by grammatical inference'', in preparation.

Albert, R., Jeong, H. \& Barab\'{a}si, A. [2000] ``Error and attack
tolerance of complex networks'', \textit{Nature} 406, 378-382.

Albert, R. \& Barab\'{a}si, A.-L. [2002] ``Statistical mechanics of complex
networks'', \textit{Rev. Mod. Phys.} 74, 47-97.

Aldridge, J. \& Pye, E. K. [1976] ``Cell density dependence of oscillatory
mechanisms'', \textit{Nature} 259, 670-671.

Amiet, J.-P. \& Huguenin, P. [1980] ``Generating functions of canonical
maps'', \textit{Helvetica Physica Acta} 53, 377-397.

Ara\'{u}jo, T. \& Vilela Mendes, R. [2000] ``Function and form in networks
of interacting agents, \textit{Complex Systems} 12, 357-378.

Ara\'{u}jo, T., Vilela Mendes, R. \& Seixas, J. [2003]; ``A dynamical
characterization of the small-world phase'', \textit{Phys. Lett. A }319,
285-289.

Ara\'{u}jo, V. [2000] ``Attractors and time averages for random maps'', 
\textit{Ann. Inst. H. Poincar\'{e} Anal. Non-lin\'{e}aire} 17, 307-369.

Barreira, L. [2002] ``Hyperbolicity and recurrence in dynamical systems: A
survey of recent results'', math.DS/0210267, \textit{Resenhas} 5, 171-230.

Boccaletti, S., Burguete, J., Gonz\'{a}lez-Vi\~{n}as, W., Mancini, H. \&
Valladares, D. (Eds.) [2001] ``\textit{Space-time chaos: Characterization,
control and synchronization}'' (World Scientific, Singapore)

Boldrighini, C., Bunimovich, L. A., Cosimi, G., Frigio, S. \& Pellegrinotti,
A. [2000] ``Ising-type and other transitions in one-dimensional coupled map
lattices with sign symmetry'', \textit{J. Stat. Phys.} 102, 1271-1283.

Bonatti, C., Diaz, L. J. \& Viana, M. [2003] ``Dynamics beyond uniform
hyperbolicity. A global geometric and probabilistic approach'', \textit{IMPA
preprint} A-221.

Boots, M. \& Sasaki, A. [1999] ``Small worlds and the evolution of
virulence: infection occurs locally and at a distance'', \textit{Proc. Roy.
Soc. London B} 266, 1933-1938.

Brutlag, D. L., Galper, A. R. \& Mills, D. H. [1991] ``Knowledge-based
simulation of DNA metabolism: Prediction of enzime action'', \textit{%
Computer Applications in the Biosciences} 7, 9-19.

Buck, J. [1988] ``Synchronous rhythmic flashing of fireflies'', \textit{%
Quart. Rev. Bio.} 63, 265-289.

Bunimovich, L. A. \& Sinai, Ya. [1988] ``Spacetime chaos in coupled map
lattices'', \textit{Nonlinearity} 1, 491-516.

Carreira, A., Hongler, M. O. \& Vilela Mendes, R. [1991] ``Variational
formulation and ergodic invariants'', \textit{Phys. Lett. A }155, 388-396.

Carvalho, R., Fernandez, B. \& Vilela Mendes, R. [2001] ``From
synchronization to multistability in two coupled quadratic maps'', \textit{%
Phys. Lett. A }285, 327-338.

Chua, L. O. \& Yang, L. [1988a] ``Cellular neural networks: Theory'', 
\textit{IEEE Trans. Circuits Syst.} 35, 1257-1272.

Chua, L. O. \& Yang, L. [1988b] ``Cellular neural networks: Applications'', 
\textit{IEEE Trans. Circuits Syst.} 35, 1273-1290.

Cohen, M. A. \& Grossberg, S. [1983] ``Absolute stability of global pattern
formation and parallel memory storage by competitive neural networks'', 
\textit{IEEE Transactions on Syst., Man and Cybern.} 13, 815-826.

Cohen, M. A. [1992] ``The construction of arbitrary stable dynamics in
nonlinear neural networks'', \textit{Neural Networks} 5, 83-103.

Colli, E. [1998] ``Infinitely many coexisting strange attractors'', \textit{%
Ann. Inst. H. Poincar\'{e} Anal. Non-lin\'{e}aire} 15, 539-579.

Coutinho, R. \& Fernandez, B. [1997a] ``Extended symbolic dynamics in
bistable coupled map lattices: Existence and stability of fronts'', \textit{%
Physica D} 108, 60-80.

Coutinho, R. \& Fernandez, B. [1997b] ``On the global orbits in a bistable
coupled map lattice'', \textit{Chaos} 7, 301-310.

Crehan, P. [1994] ``On the local Hamiltonian structure of vector fields'', 
\textit{Mod. Phys. Lett. A }9, 1399-1405.

Cross, M. C. \& Hohenberg, P. C. [1993] ``Pattern formation outside of
equilibrium'', \textit{Rev. Mod. Phys.} 65, 851-1112.

Crutchfield, J. P. [1994] ``The calculi of emergence: Computation, dynamics
and induction'', \textit{Physica D} 75, 11-54.

Crutchfield, J. P. \& Young, K. [1989] ``Inferring statistical complexity'', 
\textit{Phys. Rev. Lett.} 63, 105-108.

Crutchfield, J. P. \& Feldman, D. P. [1997] ``Statistical complexity of
simple 1D spin systems'', \textit{Phys. Rev. E} 55, 1239R-1243R.

de Jong, H. [2002] ``Modeling and simulations of genetic regulatory systems:
A literature review'', \textit{J. Comp. Biology} 9, 67-103.

Dorogovtsev, S. N. \& Mendes, J. F. [2003] ``\textit{Evolution of networks:
From biological nets to the internet and WWW}'' (Oxford Univ. Press, Oxford)

Doyne Farmer, J. [1990] ``A Rosetta stone for connectionism'', \textit{%
Physica D} 42, 153-187.

Duarte, J. T. \& Vilela Mendes, R. [1983] ``Deformation of Hamiltonian
dynamics and constants of motion in dissipative systems'', \textit{J. Math.
Phys.} 24, 1772-1778.

Dutta, M., Nusse, H. E., Ott, E., Yorke, J. A. \& Yuan, G. [1999] ``Multiple
attractor bifurcations: A source of unpredictability in piecewise smooth
systems'', \textit{Phys. Rev. Lett.} 83, 4281-4284.

Fernandez, B. \& Guiraud, P. [2004] ``Route to chaotic synchronization in
coupled map lattices: Rigorous results'', \textit{Discrete Contin. Dyn.
Systems} - Ser. B 4, 435-455.

Feudel, U., Grebogi, C., Hunt, B. R. \& Yorke, J. A. [1996] ``Maps with more
than 100 coexisting low-period periodic attractors'', \textit{Phys. Rev. E }%
54, 71-81.

Feudel, U. \& Grebogi, C. [1997] ``Multistability and the control of
complexity'', \textit{Chaos} 7, 597-604.

Fiedler, B. \& Gedeon, T. [1998] ``A class of convergent neural network
dynamics'', \textit{Physica D }111, 288-294.

Fogelman Souli\'{e}, F., Mejia, C., Goles, E. \& Martinez, S. [1989]
``Energy functions in neural networks with continuous local functions'', 
\textit{Complex Systems} 3, 269-293.

Freeman, W. J. [1992] ``Tutorial in Neurobiology: From single neurons to
brain chaos'', \textit{Int. J. of Bifurcation and Chaos} 2, 451-482.

Freidlin, M. I. \& Wentzell, A. D. [1984] ``\textit{Random perturbations of
dynamical systems''} (Springer-Verlag, New York)

Freidlin, M. I. \& Wentzell, A. D. [1994] ``\textit{Random perturbations of
Hamiltonian systems''}, Memoirs of the American Mathematical Society no. 523.

Gade, P. M. \& Hu, C.-K. [2000]; ``Synchronous chaos in coupled map lattices
with small-world interactions'', \textit{Phys. Rev. E} 62, 6409-6413.

Gambaudo, J. M. \& Tresser, C. [1983] ``Simple models for bifurcations
creating horseshoes'', \textit{J. Stat. Phys.} 32, 455-476.

Gielis, G. \& MacKay, R. S. [2000] ``Coupled map lattices with phase
transition'' \textit{Nonlinearity} 13, 867-888.

Glass, L. [1975] ``Classification of biological networks by their
qualitative dynamics'', \textit{J. Theor. Biology} 54, 85-107.

Golomb, D. \& Hansel, D. [2000]\ ``The number of synaptic inputs and the
synchrony of large sparse neuronal networks'', \textit{Neural Comput.} 12,
1095-1139.

Gouz\'{e}, J.-L. [1998] ``Positive and negative circuits in dynamical
systems'', \textit{J. of Biological Systems} 6, 11-15.

Grossberg, S. [1988] ``Nonlinear neural networks: Principles, mechanisms and
architectures'', \textit{Neural Networks} 1, 17-61.

Hanson, J. E. \& Crutchfield, J. P. [1997] ``Computational mechanics of
cellular automata: An example'', \textit{Physica D} 103, 169-189.

Henson, S. M. [2000] ``Multiple attractors and resonance in periodically
forced population models'', \textit{Physica D }140, 33-49.

Hofest\"{a}dt, R. \& Meineke, F. [1995] ``Interactive modeling and
simulation of biochemical networks'', \textit{Computers in Biology and
Medicine} 25, 321-334.

Hopfield, J. J. [1984] ``Neurons with graded response have collective
properties like those of two-state neurons'', \textit{Proc. Nat. Acad.
Sciences USA} 81, 3088-3092.

Hoppensteadt, F. C. \& Izhikevich, E. M. [1997] ``\textit{Weakly coupled
neural networks}'' (Springer, NY)

Howse, J. W., Abdallah, C. T. \& Heileman, G. L. [1996] ``Gradient and
Hamiltonian dynamics applied to learning in neural networks'', in \textit{%
Advances in Neural Information Processing Systems 8} (MIT Press)

Jiang, Z. \& McCall, M. [1993] ``Numerical simulation of a large number of
coupled lasers'', \textit{J. Opt. Soc. Am.} 10, 155.

Jiang, M. \& Pesin, Ya. [1998] ``Equilibrium measures for coupled map
lattices: Existence, uniqueness and finite-dimensional approximations'', 
\textit{Comm. Math. Phys.} 193, 675-711.

Kaneko, K. (Ed.) [1993] ``\textit{Theory and applications of coupled map
lattices'' } (Wiley, London)

Kaneko, K. [1997] ``Dominance of Milnor attractors and noise-induced
selection in a multiattractor system'', \textit{Phys. Rev. Lett.} 78,
2736-2739.

Kaneko, K. \& Tsuda, I. [2000] ``\textit{Chaos and beyond''} (Springer,
Berlin)

Katok, A. \& Hasselblatt, B.\ [1995] ``\textit{Introduction to the modern
theory of dynamical systems''} (Encyclopedia of Mathematics and its
Applications 54, Cambridge Univ. Press)

Kauffman, S. A. [1993] ``\textit{The origins of order: Self-Organization and
selection in evolution}'' (Oxford Univ. Press, New York)

Keeling, M. J. [1999] ``The effects of local spatial structure on
epidemiological invansions'', \textit{Proc. Roy. Soc. London B} 266, 859-867.

Kifer, Yu. I. [1974] ``On small random perturbations of some smooth
dynamical systems'', \textit{Mat. USSR-Izv.} 8, 1083-1107.

Kohn, K. W. [1999] ``Molecular interaction map of the mammalian cell cycle
control and DNA repair systems'', \textit{Molecular Biol. of the Cell} 10,
2703-2734.

Kulkarni, R. V., Almaas, E. \& Stroud, D. [1999] ``Evolutionary dynamics in
the Bak-Sneppen model on small-world networks'', cond-mat/9905066.

Kuramoto, Y. [1984] ``\textit{Chemical Oscillations, Waves and Turbulence''}
(Springer, Berlin)

Kuramoto, Y. [1991] ``Collective synchronization of pulse-coupled
oscillators and excitable units'', \textit{Physica D} 50, 15-30.

Lago-Fernandez, L. F., Huerta, R., Corbacho, F. \& Siguenza, J. [2000]
``Fast response and temporal coherent oscillations in small world
networks'', \textit{Phys. Rev. Lett.} 84, 2758-2761.

Laurent, M. \& Kellershohn, N. [1999] ``Multistability: a major means of
differentiation and evolution in biological systems'', \textit{Trends in
Biochem. Sci.} 24, 418-422.

Lewis, J. E. \& Glass, L. [1991] ``Steady states, limit cycles and chaos in
models of complex biological networks'', \textit{Int. J. Bifurcation and
Chaos} 1, 477-483.

Lima, R. \& Vilela Mendes, R. [1988] ``Stability of invariant circles in a
class of dissipative maps'', \textit{Nonlinear Analysis} 12, 1061-1067.

Lloyd, A. L. \& May, R. M. [2001] ``How virus spread among computers and
people'', \textit{Science} 292, 1316-1317.

MacKay, R. S. [1994] ``Mode-locking and rotational chaos in networks of
oscillators: A mathematical framework'', \textit{J. Nonlinear Sci.} 4,
301-314.

Martins, J. F., Dente, J. A., Pires, A. J. \& Vilela Mendes, R. [2001]
``Language identification of controlled systems: Modeling, control and
anomaly detection'', \textit{IEEE Trans. on Syst. Man and Cybernetics} 31,
234-242.

Martins, J.F. \& Vilela Mendes, R. [2001] ``Neural networks and logical
reasoning systems: A translation table'', \textit{Int. Journal of Neural
Systems} 11, 179-186.

May, P. \& May, E. [1999] ``Twenty years of p53 research: Structural and
functional aspects of the p53 protein'', \textit{Oncogene} 18, 7621-7636.

Meyers, S. \& Friedland, P. [1984] ``Knowledge-based simulation of genetic
regulation in bacteriophage lambda'', \textit{Nucleic Acids Research} 12,
1-9.

Newhouse, S. [1970] ``Nondensity of Axiom A(a) on $S^{2}$ '', \textit{Proc.
A. M. S. Symp. Pure Mat.} 14, 191-202.

Newhouse, S. [1974] ``Diffeomorphisms with infinitely many sinks'', \textit{%
Topology} 13, 9-18.

Newhouse, S. [1979] ``The abundance of wild hyperbolic sets and non-smooth
stable sets for diffeomorphisms'', \textit{Publ. Math. I.H.E.S.} 50, 101-151.

Palis, J. \& Viana, M. [1994] ``High dimension diffeomorphisms displaying
infinitely many periodic attractors'', \textit{Annals of Math.} 140, 207-250.

Pastor-Satorras, R. \& Vespignani, A. [2002] ``Immunisation of complex
networks'', \textit{Phys. Rev. E }65, 036104.

Pastor-Satorras, R., Rubi, M. \& Diaz-Guilera, A. [2003] (Eds.); ``\textit{%
Statistical mechanics of complex networks''} (Springer, Berlin)

Pecora, L. M. \& Carroll, T. L. [1990,1991] ``Synchronization in chaotic
systems''\textit{,} \textit{Phys. Rev. Lett.} 64 (1990) 821-824; ``Driving
systems with chaotic signals'', \textit{Phys. Rev. A }44 (1991) 2374-2383.

Pecora, L. M., Carrol, T. L. \& Heagy, J. F. [1999] ``Synchronization in
chaotic systems, Concepts and applications'' in \textit{Handbook of Chaos
Control}, ch. 10, H. G. Schuster (Ed.), (Wiley-VCH, Weinheim)

Pesin, Ya. [1997] ``\textit{Dimension theory in dynamical systems:
Contemporary views and applications'' (}Chicago Lectures in Mathematics,
Chicago Univ. Press)

Peskin, C. S. [1975] ``\textit{Mathematical Aspects of Heart Physiology''},
Courant Inst. Math. Sciences, New York.

Pikovsky, A., Rosenblum, M. \& Kurths, J. [2001]; ``\textit{Synchronization.
A universal concept in nonlinear sciences''} (Cambridge Univ. Press)

Plahte, E., Mestl, T. \& Omholt, S. W. [1995] ``Stationary states in food
web models with threshold relationships'', \textit{J. of Biological Systems}
3, 569-577.

Pontryagin, L. S. [1934] ``On dynamical systems close to Hamiltonian
systems'' (Russian), \textit{Zh. Eksp. Teor. Fiz.} 4, 234-238.

Poon, L. \& Grebogi, C. [1995] ``Controlling complexity'', \textit{Phys.
Rev. Lett.} 75, 4023-4026.

Robinson, C. [1983] ``Bifurcation to infinitely many sinks'', \textit{Comm.
Math Phys.} 90, 433-459.

Roels, J. [1974] ``On the local decomposition of a vector field on a
symplectic surface as the sum of a gradient and a Hamiltonian field'' 
\textit{Comptes Rendus Acad. Sci. Paris} 278 SerA., 29-31.

Ruelle, D. [1987] ``Theory and experiment in the ergodic study of chaos and
strange attractors'' in \textit{Proc. VIII Int. Congress on Mathematical
Physics}, pgs. 273-282, M. Mebkhout and R. S\'{e}n\'{e}or (Eds.) (World
Scientific, Singapore)

Sch\"{u}rmann, B. [1989] ``Stability and adaptation in artificial neural
systems'', \textit{Phys. Rev. A }40, 2681-2688.

Shalizi, C. R. \& Crutchfield, J. P. [2001] ``Computational mechanics:
Pattern and prediction, structure and simplicity'', \textit{J. Stat. Phys.}
104, 817-879.

Sharpless, N. E. \& DePinho, R. A. [2002] ``p53: Good cop/Bad cop'', \textit{%
Cell} 110, 9-12.

Shimada, T., Hagiya, M., Arita, M., Nishizaki, S. \& Tan, C. L. [1995]
``Knowledge-based simulation of regulatory action in lambda phage'', \textit{%
Int. J. of Artificial Intelligence Tools }4, 511-524.

Skarda, C. A. \& Freeman, W. J. [1987]\ ``How brains make chaos in order to
make sense of the world'', \textit{Behavioral and Brain Sciences} 10,
161-195.

Snoussi, E. H. \& Thomas, R. [1993] ``Logical identification of all steady
states: The concept of feedback loop-characteristic state'', \textit{Bull.
Math. Biology} 55, 973-991.

Snoussi, E. H. [1998] ``Necessary conditions for multistationarity and
stable periodicity'', \textit{J. of Biological Systems} 6, 3-9.

Somogyi, R. \& Sniegoski, C. A. [1996] ``Modeling the complexity of genetic
networks: Understanding multigenic and pleiotropic regulation'', \textit{%
Complexity} 1, 45-63.

Strogatz, S. H. [2000] ``From Kuramoto to Crawford: exploring the onset of
synchronization in populations of coupled oscillators'', \textit{Physica D }%
143, 1-20.

Strogatz, S. H. [2001] ``Exploring complex networks, \textit{Nature} 410,
268-276.

Taylor, W. R. et al. [1999] ``Mechanisms of G2 arrest in response to
overexpression of p53'', \textit{Mol. Biol. of the Cell} 10, 3607-3622.

Terman, D. [2002] ``Dynamics of singularly perturbed neuronal networks'' in 
\textit{An Introduction to Mathematical Modeling in Physiology, Cell Biology
and Immunology}, Proc. Symp. in Applied Math. vol. 59, pages 1-32, Am. Math.
Soc., Providence 2002.

Thieffry, D., Colet, M. \& Thomas, R. [1993]; ``Formalization of regulatory
networks: A logical method and its automatization'', \textit{Math. Modeling
Sci. Computing} 2, 144-151.

Thomas, R. \& D'Ari [1990] \textit{``Biological Feedback''} (CRC Press, Boca
Raton)

Thomas, R. [1991] ``Regulatory networks seen as asynchronous automata: A
logical description'', \textit{J. Theor. Biology} 153, 1-23.

Thomas, R., Thieffry, D. \& Kaufman, M. [1995] ``Dynamical behavior of
biological regulatory networks - I. Biological role of feedback loops and
practical use of the concept of loop-characteristic state'', \textit{Bull of
Math. Biology} 57, 247-276.

Tonti, E. [1992] ``Variational formulation for every nonlinear equation'',
Trieste ICTP report SMR/92-20 and references therein.

Tyson, J. J. \& Ohtmer, H. G. [1978] ``The dynamics of feedback control
circuits in biochemical pathways'', \textit{Progress in Theor. Biology} 5,
1-62.

Vilela Mendes, R. \& Duarte, J. T. [1981] ``Decomposition of vector fields
and mixed dynamics'', \textit{J. Math. Phys.} 22, 1420-1422.

Vilela Mendes, R. \& Duarte, J. T. [1982] ``Arcs of discrete dynamics and
constants of motion'', \textit{Lett. Math. Phys.} 6, 249-252.

Vilela Mendes, R. [1984] ``A variational formulation for dissipative maps'', 
\textit{Phys. Lett. A }104, 391-395.

Vilela Mendes, R. [1986] ``Generating functions for noncanonical maps'', 
\textit{Lett. Math. Phys.} 11, 289-292.

Vilela Mendes, R. [1988] ``Deformation stability of periodic and
quasiperiodic motion in dissipative systems'' in \textit{Deformation theory
of algebras and structures and applications}, M. Hazewinkel and M.
Gerstenhaber (Eds.) (Kluwer Acad. Pub., Dordrecht)

Vilela Mendes, R. \& Duarte, J. T. [1992] ``Vector fields and neural
networks'', \textit{Complex Systems} 6, 21-30.

Vilela Mendes, R. [1998] ``Conditional exponents, entropies and a measure of
dynamical self-organization'', \textit{Phys. Lett. A }248, 167 - 171.

Vilela Mendes, R. [1999] ``Clustering and synchronization with positive
Lyapunov exponents'', \textit{Phys. Lett. A }257, 132-138.

Vilela Mendes, R. [2000a] ``Multistability in dynamical systems'' in ''%
\textit{Dynamical systems: From crystal to chaos}'', J.-M.Gambaudo, P.
Hubert, P. Tisseur and S. Vaienti (Eds.), pp. 105-113 (World Scientific,
Singapore)

Vilela Mendes, R. [2000b] ``Characterizing self-organization and coevolution
by ergodic invariants'', \textit{Physica A }276, 550-571.

Vilela Mendes, R. [2001] ``Structure generating mechanisms in agent-based
models'', \textit{Physica A} 295, 537-561.

Vogelstein, B., Lane, D. \& Levine, A. J. [2000] ``Surfing the p53
network'', \textit{Nature} 408, 307-310.

Vousden, K. H. [2000] ``p53: Death star'', \textit{Cell} 103, 691-694.

Wang, X.-J. [1990] ``The Newhouse set has a positive Hausdorff dimension'', 
\textit{Comm. Math. Phys.} 131, 317-332.

Watts, D. J. \& Strogatz, S. H. [1998] ``Collective dynamics of small-world
networks'', \textit{Nature} 393, 440-442.

Watts, D. J. [1999] ``\textit{Small Worlds: The dynamics of networks between
order and randomness''} (Priceton Univ. Press, Princeton NJ)

Wei, G. W., Zhan, M. \& Lai, C.-H. [2002] ``Tayloring wavelets for chaos
control'', \textit{Phys. Rev. Lett.} 89, 284103.

Weigel, R. \& Atlee Jackson, E. [1998] ``Experimental implementation of
migrations in multiple attractor systems'', \textit{Int. J. of Bifurcation
and Chaos} 8, 173-178.

Winfree, A. T. [1967]\ ``Biological rhytms and the behavior of populations
of coupled oscillators'', \textit{J. Theor. Bio.} 16, 15-42.

Woods, D. B. \& Vousden, K. H. [2001] ``Regulation of p53 function'', 
\textit{Exp. Cell Reserach} 264, 56-66.

Wuensche, A. [2002] ``Basins of attraction in network dynamics: A conceptual
framework for biomolecular networks'', SFI Working paper 02-02-004 to appear
in \textit{Modularity in Development and Evolution}, G. Schlosser and G. P.
Wagner (Eds.) (Chicago Univ. Press.)

Yang, X.-S. [2001] ``Chaos in small-world networks'', \textit{Phys. Rev. E}
63, 046206.

\end{document}